# All-Optical Synthesis of an Arbitrary Linear Transformation Using Diffractive Surfaces

Onur Kulce[1,2,3], Deniz Mengu[1,2,3], Yair Rivenson[1,2,3], Aydogan Ozcan[1,2,3,*]

[1] Electrical and Computer Engineering Department, University of California, Los Angeles, CA, 90095, USA

[2] Bioengineering Department, University of California, Los Angeles, CA, 90095, USA

[3] California NanoSystems Institute, University of California, Los Angeles, CA, 90095, USA

* Corresponding author: ozcan@ucla.edu

*Onur Kulce*: onurkulce@ucla.edu

*Deniz Mengu*: denizmengu@ucla.edu

*Yair Rivenson*: rivensonyair@ucla.edu

*Aydogan Ozcan*: ozcan@ucla.edu

Telephone: +1 310-825-0915

Postal Address: 420 Westwood Plaza, UCLA, Los Angeles, CA, USA, 90095

**Abstract**

Spatially-engineered diffractive surfaces have emerged as a powerful framework to control light-matter interactions for e.g., statistical inference and the design of task-specific optical components. Here, we report the design of diffractive surfaces to all-optically perform arbitrary complex-valued linear transformations between an input $(N_i)$ and output $(N_o)$, where $N_i$ and $N_o$ represent the number of pixels at the input and output fields-of-view (FOVs), respectively. First, we consider a single diffractive surface and use a matrix pseudoinverse-based method to determine the complex-valued transmission coefficients of the diffractive features/neurons to all-optically perform a desired/target linear transformation. In addition to this *data-free* design approach, we also consider a deep learning-based design method to optimize the transmission coefficients of diffractive surfaces by using examples of input/output fields corresponding to the target transformation. We compared the all-optical transformation errors and diffraction efficiencies achieved using data-free designs as well as data-driven (deep learning-based) diffractive designs to all-optically perform (*i*) arbitrarily-chosen complex-valued transformations including unitary, nonunitary and noninvertible transforms, (*ii*) 2D discrete Fourier transformation, (*iii*) arbitrary 2D permutation operations, and (*iv*) high-pass filtered coherent imaging. Our analyses reveal that if the total number $(N)$ of spatially-engineered diffractive features/neurons is $\geq N_i \times N_o$, both design methods succeed in all-optical implementation of the target transformation, achieving negligible error. However, compared to data-free designs, deep learning-based diffractive designs are found to achieve significantly larger diffraction efficiencies for a given $N$ and their all-optical transformations are more accurate for $N < N_i \times N_o$. These conclusions are generally applicable to various optical processors that employ spatially-engineered diffractive surfaces.

## 1. Introduction

It is well-known that optical waves can be utilized for the processing of spatial and/or temporal information[1–6]. Using optical waves to process information is appealing since computation can be performed at the speed of light, with high parallelization and throughput, also providing potential power advantages. For this broad goal, various optical computing architectures have been demonstrated in the literature[7–21]. With the recent advances in photonic material engineering, e.g., metamaterials, metasurfaces and plasmonics, the utilization of advanced diffractive materials that can precisely shape optical wavefronts through light-matter interaction has become feasible[22–27]. For example, optical processors formed through spatially-engineered diffractive surfaces have been shown to achieve both statistical inference and deterministic tasks, such as image classification, single-pixel machine vision and spatially-controlled wavelength division multiplexing, among others[28–37].

Since scalar optical wave propagation in free space and light transmission through diffractive surfaces constitute linear phenomena, the light transmission from an input field-of-view (FOV) to an output FOV that is engineered through diffractive surfaces can be formulated using linear algebra[35]. As a result, together with the free space diffraction, the light transmittance patterns of diffractive surfaces (forming an optical network) collectively define a certain complex-valued all-optical linear transformation between the input and output FOVs. In this paper, we focus on designing these spatial patterns and diffractive surfaces that can all-optically compute a desired, target transformation. We demonstrate that an arbitrary complex-valued linear transformation between an input and output FOV can be realized using spatially-engineered diffractive surfaces, where each feature (neuron) of a diffractive layer modulates the amplitude and/or phase of the optical wave field. In generating the needed diffractive surfaces to all-optically achieve a given target transformation, we use both a matrix pseudoinverse-based design that is *data-free* as well as a data-driven, deep learning-based design method. In our analysis, we compared the approximation capabilities of diffractive surfaces for performing various all-optical linear transformations as a function of the total number of diffractive neurons, number of diffractive layers and the area of the input/output FOVs. For these comparisons, we used as our target transformations arbitrarily generated complex-valued unitary, nonunitary and noninvertible transforms, 2D Fourier transform, 2D random permutation operation as well as high-pass filtered coherent imaging operations.

Our results reveal that when the total number of engineered/optimized diffractive neurons of a material design exceeds $N_i \times N_o$, both the data-free and data-driven diffractive designs successfully approximate the target linear transformation with negligible error; here, $N_i$ and $N_o$ refer to the number of diffraction-limited, independent spots/pixels located within the area of the input and output FOVs, respectively. This means, to all-optically perform an arbitrary complex-valued linear transformation between larger input and/or larger output FOVs, larger area diffractive layers with more neurons or a larger number of diffractive layers need to be utilized. Our analyses further reveal that deep learning-based data driven diffractive designs (that learn a target linear transformation through examples of input/output fields) overall achieve much better diffraction efficiency at the output FOV. All in all, our analysis confirms that for a given diffractive layer size, with a certain number of diffractive features per layer (like a building block of a diffractive network), the creation of deeper diffractive networks with one layer following another, can improve both the transformation error and the diffraction efficiency of the resulting all-optical transformation.

Our results and conclusions can be broadly applied to any part of the electromagnetic spectrum to design all-optical processors using spatially-engineered diffractive surfaces to perform an arbitrary complex-valued linear transformation.

## 2. Results

### 2.1. Formulation of all-optical transformations using diffractive surfaces

Let $\boldsymbol{i}$ and $\boldsymbol{o}$ be the column vectors that include the samples of the 2D complex-valued input and output FOVs, respectively, as shown in Fig. 1.a. Here we assume that the optical wave field can be represented using the scalar field formulation[38–40]. $\boldsymbol{i}$ and $\boldsymbol{o}$ are generated by, first, sampling the 2D input and output FOVs, and then vectorizing the resulting 2D matrices in a column-major order. Following our earlier notation, $N_i$ and $N_o$ represent the number of diffraction-limited spots/pixels on the input and output FOVs, respectively, which also define the lengths of the vectors $\boldsymbol{i}$ and $\boldsymbol{o}$. In our simulations, we assume that the sampling period along both the horizontal and vertical directions is $\lambda/2$, where $\lambda$ is the wavelength of the monochromatic scalar optical field. With this selection in our model, we include all the propagating modes that are transmitted through the diffractive layer(s).

To implement the wave propagation between parallel planes in free space, we generate a matrix, $\boldsymbol{H}_d$, where $d$ is the axial distance between two planes (e.g., $d \geq \lambda$). Since this matrix represents a convolution operation where the 2D impulse response originates from the Rayleigh-Sommerfeld diffraction formulation[3], it is a Toeplitz matrix[41]. We generate this matrix using the Fourier relation in the discrete domain as

$$\boldsymbol{H}_d = \boldsymbol{W}^{-\boldsymbol{1}}\boldsymbol{D}\boldsymbol{W} = \boldsymbol{W}^H\boldsymbol{D}\boldsymbol{W} \qquad 1$$

where $\boldsymbol{W}$ and $\boldsymbol{W}^{-\boldsymbol{1}}$ are the 2D discrete Fourier transform (DFT) and inverse discrete Fourier transform (IDFT) matrices, respectively, and the superscript $H$ represents the matrix Hermitian operation. We choose the scaling constant appropriately such that the unitarity of the DFT operation is preserved, i.e., $\boldsymbol{W}^{-\boldsymbol{1}} = \boldsymbol{W}^H$ [41]. The matrix, $\boldsymbol{D}$, represents the transfer function of free space propagation in the 2D Fourier domain and it includes nonzero elements only along its main diagonal entries. These entries are the samples of the function, $e^{jk_z d}$, for $0 \leq k_z = \sqrt{k^2 - \left(k_x^2 + k_y^2\right)} \leq k$, where $k_x, k_y \in [-k, k)$. Here $k = 2\pi/\lambda$ is the wavenumber of the monochromatic optical field and $\left(k_x, k_y\right)$ pair represents the 2D spatial frequency variables along $x$ and $y$ directions, respectively[3]. To ignore the evanescent modes, we choose the diagonal entries of $\boldsymbol{D}$ that correspond to the $k_z$ values for $k^2 \leq k_x^2 + k_y^2$ as zero; since $d \geq \lambda$ this is an appropriate selection. In our model, we choose the 2D discrete wave propagation square window size, $\sqrt{N_d} \times \sqrt{N_d}$, large enough (e.g., $N_d = 144^2$) such that the physical wave propagation between the input plane, diffractive layers and the output plane is simulated accurately [42]. Also, since $\boldsymbol{H}_d$ represents a convolution in 2D space, the entries of $\boldsymbol{W}$, $\boldsymbol{W^{-1}}$ and $\boldsymbol{D}$ follow the same vectorization procedure applied to the input and output FOVs. As a result, the sizes of all these matrices become $N_d \times N_d$.

Since the diffractive surfaces are modeled as thin elements, the light transmission through surfaces can be formulated as a pointwise multiplication operation, where the output optical field of a layer equals to its input optical field multiplied by the complex-valued transmission function, $t(x, y)$, of the diffractive layer. Hence, in matrix formulation, this is represented by a diagonal matrix $\boldsymbol{T}$, where the diagonal entries are the vectorized samples of $t(x, y)$. Hence the size of $\boldsymbol{T}$ becomes $N_L \times N_L$, where $N_L$ is the total number of diffractive features (referred to as neurons) on the corresponding layer.

We also assume that the forward propagating optical fields are zero outside of the input FOV and outside of the transmissive parts of each diffractive surface, so that we solely analyze the modes that are propagating through the transmissive diffractive layers. This is not a restrictive assumption as it can be simply satisfied by introducing light blocking, opaque materials around the input FOV and the diffractive layers. Furthermore, although the wave field is not zero outside of the output FOV, we only formulate the optical wave propagation between the input and output FOVs since the complex-valued transformation ($\boldsymbol{A}$) that we would like to approximate is *defined* between $\boldsymbol{i}$ and $\boldsymbol{o}$. As a result of these, we delete the appropriate rows and columns of $\boldsymbol{H}_d$, which are generated based on Equation 1. We denote the resulting matrix as $\boldsymbol{H}'_d$.

Based on these definitions, the relationship between the input and output FOVs for a diffractive network that has $K$ diffractive layers can be written as

$$\boldsymbol{o}' = \boldsymbol{H}'_{d_{K+1}} \boldsymbol{T}_K \boldsymbol{H}'_{d_K} \cdots \boldsymbol{T}_2 \boldsymbol{H}'_{d_2} \boldsymbol{T}_1 \boldsymbol{H}'_{d_1} \boldsymbol{i} = \boldsymbol{A}' \boldsymbol{i} \qquad 2$$

as shown in Fig. 1.a. Here $d_1$ is the axial distance between the input FOV and the first diffractive layer, $d_{K+1}$ is the axial distance between the $K^{th}$ layer and the output FOV, and $d_l$ for $l \in \{2,3,\cdots,K\}$ is the axial distance between the $(l-1)^{th}$ and $l^{th}$ diffractive layers (see Fig. 1.a). Also, $\boldsymbol{T}_l$ for $l \in \{1,2, \cdots, K\}$ is the complex-valued light transmission matrix of the $l^{th}$ layer. The size of $\boldsymbol{H}'_{d_1}$ is $N_{L_1} \times N_i$, the size of $\boldsymbol{H}'_{d_{K+1}}$ is $N_o \times N_{L_K}$ and the size of $\boldsymbol{H}'_{d_l}$ is $N_{L_l} \times N_{L_{l-1}}$ for $l \in \{2,3, \cdots, K\}$, where $N_{L_l}$ is the number of diffractive neurons at the $l^{th}$ diffractive layer. Note that, in our notation in Equation 2, we define $\boldsymbol{o}'$ as the calculated output by the diffractive system, whereas $\boldsymbol{o}$ refers to the ground truth/target output in response to $\boldsymbol{i}$. The matrix $\boldsymbol{A}'$ in Equation 2, that is formed by successive diffractive layers/surfaces, represents the all-optical transformation performed by the diffractive network from the input FOV to the output FOV. Note that this formalism does not aim to optimize the diffractive system in order to implement only one given pair of input-output complex fields; instead it aims to all-optically approximate an arbitrary complex-valued linear transformation, $\boldsymbol{A}$.

## 2.2. Matrix pseudoinverse-based synthesis of an arbitrary complex-valued linear transformation using a single diffractive surface $(\boldsymbol{K} = \boldsymbol{1})$

In this section, we focus on data-free design of a single diffractive layer ($K = 1$), in order to determine the diagonal entries of $\boldsymbol{T}_1$ such that the resulting transformation matrix, $\boldsymbol{A}'$ which is given by Equation 2, approximates the transformation matrix $\boldsymbol{A}$. To accomplish this, we first vectorize $\boldsymbol{A}'$ in a column-major order and write it as[35]

$$\begin{aligned} vec(\boldsymbol{A}') = \boldsymbol{a}' &= vec\left(\boldsymbol{H}'_{d_2}\boldsymbol{T}_1\boldsymbol{H}'_{d_1}\right) \\ &= \left(\boldsymbol{H}'^{T}_{d_1} \otimes \boldsymbol{H}'_{d_2}\right) vec(\boldsymbol{T}_1) \end{aligned} \quad (3)$$

where $\otimes$ and the superscript $T$ represent the Kronecker product and the matrix transpose operator, respectively. Since the elements of $vec(\boldsymbol{T}_1)$ are nonzero only for the diagonal elements of $\boldsymbol{T}_1$, Equation 3 can be further simplified as

$$\boldsymbol{a}' = \boldsymbol{H}'\boldsymbol{t}_1 \quad (4)$$

where $\boldsymbol{t}_1[l] = \boldsymbol{T}_1[l,l]$ and $\boldsymbol{H}'[:,l] = \boldsymbol{H}'^{T}_{d_1}[:,l] \otimes \boldsymbol{H}'_{d_2}[:,l]$ for $l \in \{1,2, \cdots, N_{L_1}\}$, and $[:,l]$ represents the $l^{th}$ column of the associated matrix in our notation. Here the matrix $\boldsymbol{H}'$ has size $N_i N_o \times N_{L_1}$ and is a full-rank matrix with rank $D = min(N_i N_o, N_{L_1})$ for $d_1 \neq d_2$. If $d_1 = d_2$, the maximum rank reduces to $N_i(N_i + 1)/2$ when $N_i = N_o$ [35]. We assume that $d_1 \neq d_2$ and denote the maximum achievable rank as $D_{max}$, which is equal to $N_i N_o$.

Based on Equation 4, the computation of the neuron transmission values of the diffractive layer that approximates a given complex-valued transformation matrix $\boldsymbol{A}$ can be reduced to an L2-norm minimization problem, where the approximation error which is subject to the minimization is [41]

$$\begin{aligned} \|\boldsymbol{a} - m\boldsymbol{a}'\|^2 = \|\boldsymbol{a} - m\boldsymbol{H}'\boldsymbol{t}_1\|^2 = \|\boldsymbol{a} - \boldsymbol{H}'\hat{\boldsymbol{t}}_1\|^2 = \|\boldsymbol{a} - \hat{\boldsymbol{a}}'\|^2 \\ = \frac{1}{N_i N_o} \sum_{l=1}^{N_i N_o} |\boldsymbol{a}[l] - m\boldsymbol{a}'[l]|^2 \\ = \frac{1}{N_i N_o} \sum_{l=1}^{N_i N_o} |\boldsymbol{a}[l] - \hat{\boldsymbol{a}}'[l]|^2 \end{aligned} \quad (5)$$

where $\boldsymbol{a}$ is the vectorized form of the target transformation matrix $\boldsymbol{A}$, i.e., $vec(\boldsymbol{A}) = \boldsymbol{a}$. We included a scalar, normalization coefficient ($m$) in Equation 5 so that the resulting difference term does not get affected by a diffraction-efficiency related scaling mismatch between $\boldsymbol{A}$ and $\boldsymbol{A}'$; also note that we assume a passive diffractive layer without any optical gain, i.e., $|\boldsymbol{t}_1[l]| \leq 1$ for all $l \in \{1,2, \cdots, N_{L_1}\}$. As a result of this, we also introduced in Equation 5, $m\boldsymbol{t}_1 = \hat{\boldsymbol{t}}_1$.

Throughout the paper, we use $\left\|\boldsymbol{A} - \hat{\boldsymbol{A}}'\right\|^2$ and $\|\boldsymbol{a} - \hat{\boldsymbol{a}}'\|^2$ interchangeably both referring to Equation 5 and define them as the ***all-optical transformation error***. We also refer to $\boldsymbol{a}$, $\boldsymbol{a}'$ and $\hat{\boldsymbol{a}}'$ as the target transformation (ground truth), estimate transformation and normalized estimate transformation vectors, respectively.

If $N_{L_1} > N_i N_o$, the number of equations in Equation 4 becomes less than the number of unknowns and the matrix-vector equation corresponds to an underdetermined system. If $N_{L_1} < N_i N_o$, on the other hand, the equation system becomes an overdetermined system. In the critical case, where $N_{L_1} = N_i N_o$, $\boldsymbol{H}'$ becomes a full-rank square matrix, hence, is an invertible matrix. There are various numerical methods for solving the formulated matrix-vector equation and minimizing the transformation error given in Equation 5 [41]. In this paper, we adopt the pseudoinverse-based method among other numerical methods in computing the neuron transmission values in finding the estimate transformation $\boldsymbol{A}'$ for all the cases, i.e., $N_{L_1} > N_i N_o$, $N_{L_1} < N_i N_o$ and $N_{L_1} = N_i N_o$. For this, we compute the neuron values from a given target transformation as

$$\hat{\boldsymbol{t}}_1 = \boldsymbol{H}'^{\dagger}\boldsymbol{a} \qquad 6$$

where $\boldsymbol{H}'^{\dagger}$ is the pseudoinverse of $\boldsymbol{H}'$. This pseudoinverse operation is performed using the singular value decomposition (SVD) as

$$\boldsymbol{H}'^{\dagger} = \boldsymbol{V}\boldsymbol{\Sigma}^{-1}\boldsymbol{U}^{H} \qquad 7$$

where $\boldsymbol{U}$ and $\boldsymbol{V}$ are orthonormal matrices and $\boldsymbol{\Sigma}$ is a diagonal matrix that contains the singular values of $\boldsymbol{H}'$. $\boldsymbol{U}, \boldsymbol{V}$ and $\boldsymbol{\Sigma}$ form the $\boldsymbol{H}'$ matrix as

$$\boldsymbol{H}' = \boldsymbol{U}\boldsymbol{\Sigma}\boldsymbol{V}^{H} \qquad 8$$

To prevent the occurrence of excessively large numbers that might cause numerical artifacts, we take very small singular values as zero during the computation of $\boldsymbol{\Sigma}^{-1}$. After computing $\hat{\boldsymbol{t}}_1$, the normalization constant ($m$) and the physically realizable neuron values can be calculated as:

$$m = \max_{l}(|\hat{\boldsymbol{t}}_1|) \quad \text{and} \quad \boldsymbol{t}_1 = \hat{\boldsymbol{t}}_1/m \qquad 9$$

In summary, the vector $\boldsymbol{t}_1$ that includes the transmittance values of the diffractive layer is computed from a given, target transformation vector, $\boldsymbol{a}$, using Equations 6 and 9, and then the resulting estimate transformation vector, $\boldsymbol{a}'$, is computed using Equation 4. Finally, $\boldsymbol{A}'$, is obtained from $\boldsymbol{a}'$ by reversing the vectorization operation.

### 2.3. Deep learning-based synthesis of an arbitrary complex-valued linear transformation using diffractive surfaces $(\boldsymbol{K} \geq \mathbf{1})$

Different from the numerical pseudoinverse-based design method described in the previous section, which is data-free in its computational steps, deep learning-based design of diffractive layers utilize a training dataset containing examples of input/output fields corresponding to a target transformation $\boldsymbol{A}$. In a $K$-layered diffractive network, our optical forward model implements Equation 2, where the diagonal entries of each $\boldsymbol{T}_l$ matrix for $l \in \{1,2, \cdots, K\}$ become the arguments subject to the optimization. At each iteration of deep learning-based optimization during the error-back-propagation algorithm, the complex-valued neuron values are updated to minimize the following normalized mean-squared-error loss function:

$$\left\|\widetilde{\boldsymbol{o}}_s - \widetilde{\boldsymbol{o}}'_{s,c}\right\|^2 = \frac{1}{N_o}\sum_{l=1}^{N_o}\left|\sigma_s \boldsymbol{o}_s[l] - \sigma'_{s,c}\boldsymbol{o}'_{s,c}[l]\right|^2 \tag{10}$$

where

$$\boldsymbol{o}_s = \boldsymbol{A}\boldsymbol{i}_s \quad \text{and} \quad \boldsymbol{o}'_{s,c} = \boldsymbol{A}'_c\boldsymbol{i}_s \tag{11}$$

refer to the ground truth and the estimated output field by the diffractive network, respectively, for the $s^{th}$ input field in the training dataset, $\boldsymbol{i}_s$. The subscript $c$ indicates the current state of the all-optical transformation at a given iteration of the training that is determined by the current transmittance values of the diffractive layers. The constant $\sigma_s$ normalizes the energy of the ground truth field at the output FOV and can be written as

$$\sigma_s = \left(\sum_{l=1}^{N_o}|\boldsymbol{o}_s[l]|^2\right)^{-\frac{1}{2}} \tag{12}$$

Also, the complex valued $\sigma'_{s,c}$ is calculated to minimize Equation 10. It can be computed by taking the derivative of $\left\|\widetilde{\boldsymbol{o}}_s - \widetilde{\boldsymbol{o}}'_{s,c}\right\|^2$ with respect to $\sigma'^{*}_{s,c}$, which is the complex conjugate of $\sigma'_{s,c}$, and then equating the resulting expression to zero,[44] which yields:

$$\sigma'_{s,c} = \frac{\sum_{l=1}^{N_o} \sigma_s \boldsymbol{o}_s[l] \boldsymbol{o}'^{*}_{s,c}[l]}{\sum_{l=1}^{N_o} \left|\boldsymbol{o}'_{s,c}[l]\right|^2} \quad (13)$$

Other details of this deep learning-based, data-driven design of diffractive layers are presented in Section 4.2. After the training is over, which is a one-time effort, the estimate transformation matrix and the corresponding vectorized form, $\boldsymbol{A}'$ and $vec(\boldsymbol{A}') = \boldsymbol{a}'$, are computed using the optimized neuron transmission values in Equation 2. After computing $\boldsymbol{a}'$, we also compute the normalization constant, $m$, which minimizes $\|\boldsymbol{a} - m\boldsymbol{a}'\|^2$, resulting in:

$$m = \frac{\sum_{l=1}^{N_i N_o} \boldsymbol{a}[l] \boldsymbol{a}'^{*}[l]}{\sum_{l=1}^{N_i N_o} |\boldsymbol{a}'[l]|^2} \quad (14)$$

In summary, an optical network that includes *K* diffractive surfaces can be optimized using deep learning through training examples of input/output fields that correspond to a target transformation, $\boldsymbol{A}$. Starting with the next section, we will analyze and compare the resulting all-optical transformations that can be achieved using data-driven (deep learning-based) as well as data-free designs that we introduced.

## 2.4. Comparison of all-optical transformations performed through diffractive surfaces designed by matrix pseudoinversion vs. deep learning

In this section we present a quantitative comparison of the pseudoinverse- and deep learning-based methods in synthesizing various all-optical linear transformations between the input and output FOVs using diffractive surfaces. In our analysis, we took the total number of pixels in both the input and output FOVs as $N_i = N_o = 64$ ($i.e.$ $8 \times 8$) and the size of each $\boldsymbol{H}_d$ matrix was $144^2 \times 144^2$ with $N_d = 144^2$. The linear transformations that we used as our comparison testbeds are (*i*) arbitrarily generated complex-valued unitary transforms, (*ii*) arbitrarily generated complex-valued nonunitary and invertible transforms, (*iii*) arbitrarily generated complex-valued noninvertible transforms, (*iv*) the 2D discrete Fourier transform, (*v*) a permutation matrix-based transformation, and (*vi*) a high-pass filtered coherent imaging operation. The details of the diffractive network configurations, training image datasets, training parameters, computation of error metrics and the generation of ground truth transformation matrices are presented in Section 4. Next, we present the performance comparisons for different all-optical transformations.

### 2.4.1. Arbitrary complex-valued unitary and nonunitary transforms

In Figs. 1-3 and Supplementary Figs. S1-S3, we present the results for two different arbitrarily selected complex-valued *unitary* transforms that are approximated using diffractive surface designs with different number of diffractive layers, $K$, and different number of neurons, $N = \sum_{k=1}^{K} N_{L_k}$. Similarly, Figs. 4-6 and Supplementary Figs. S4-S6 report two different *nonunitary*, arbitrarily selected complex-valued linear transforms performed through diffractive surface designs. To cover different types of transformations, Figs. 4-6 and Supplementary Figs. S4-S6 report an invertible nonunitary and a noninvertible (hence, nonunitary) transformation, respectively. The magnitude and phase values of these target transformations ($\boldsymbol{A}$) are also shown in Figs. 1.b, 4.b, and Supplementary Figs. S1.b and S4.b.

To compare the performance of all-optical transformations that can be achieved by different diffractive designs, Fig. 1.c, Supplementary Fig. S1.c, Fig. 4.c and Supplementary Fig. S4.c report the resulting transformation errors for the above described testbeds ($\boldsymbol{A}$ matrices) as a function of $N$ and $K$. It can be seen that, in all of the diffractive designs reported in these figures, there is a monotonic decrease in the transformation error as the total number of neurons in the network increases. In data-free, matrix pseudoinverse-based designs $(K = 1)$, the transformation error curves reach a baseline, approaching $\sim 0$ starting at $N = 64^2$. This empirically-found turning point of the transformation error at $N = 64^2$ also coincides with the limit of the information processing capacity of the diffractive network dictated by $D_{max} = N_i N_o = 64^2$ [35]. Beyond this point, i.e., for $N > 64^2$, the all-optical transformation errors of data-free diffractive designs remain negligible for these complex-valued unitary as well as nonunitary transformations defined in Fig. 1.b, Supplementary Fig. S1.b, Fig. 4.b and Supplementary Fig. S4.b.

On the other hand, for data-driven, deep learning-based diffractive designs, one of the key observations is that, as the number of diffractive layers $(K)$, increases, the all-optical transformation error decreases for the same $N$. Stated differently, deep learning-based, data-driven diffractive designs prefer to distribute/divide the total number of neurons $(N)$ into different, successive layers as opposed having all the $N$ neurons at a single, large diffractive layer; the latter, deep learning-designed $K = 1$, exhibits much worse all-optical transformation error compared to e.g., $K = 4$ diffractive layers despite the fact that both of these designs have the same number of trainable neurons $(N)$. Furthermore, as illustrated in Fig. 1, Supplementary Fig. S1, Fig. 4 and Supplementary Fig. S4, deep learning-based diffractive designs with $K = 4$ layers match the transformation error performance of data-free designs based on matrix pseudoinversion and also exhibit negligible transformation error for $N \geq 64^2 = N_i N_o$. However, when $N < 64^2$ the deep learning-based diffractive designs with $K = 4$ layers achieve smaller transformation errors compared to data-free diffractive designs that have the same number of neurons. Similar conclusions can be made in Figs. 1.e, S1.e, 4.e and S4.e, by comparing the mean-squared-error (MSE) values calculated at the output FOV using test images (input fields). For $N \geq 64^2 = N_i N_o$ the deep learning-based diffractive designs $(K = 4)$ along with the data-free diffractive designs achieve output MSE values that approach $\sim 0$, similar to the all-optical transformation errors that approach $\sim 0$ in Fig. 1.c, Supplementary Fig. S4.c, Fig. 4.c and Supplementary Fig. S7.c. However for designs that have smaller number of neurons, i.e., $N < N_i N_o$, the deep learning-based diffractive designs with $K = 4$ achieve much better MSE at the output FOV compared to data-free diffractive designs that have same number of neurons $(N)$.

In addition to these, one of the most significant differences between the pseudoinverse-based data-free diffractive designs and deep learning-based counterparts is observed in the optical diffraction efficiencies calculated at the output FOV; see Figs. 1.f, 4.f, and Supplementary Figs. 1.f and S4.f. Even though the transformation errors (or the output MSE values) of the two design approaches remain the same (~0) for $N \geq 64^2 = N_i N_o$ , the diffraction efficiencies of the all-optical transformations learned using deep learning significantly outperform the diffraction efficiencies achieved using data-free, matrix pseudoinverse-based designs as shown in Figs. 1.f, S1.f, 4.f and S4.f.

On top of transformation error, output MSE and diffraction efficiency metrics, Figs. 1.d, S1.d, 4.d and S4.d also report the cosine similarity (see Section 4) between the estimated all-optical transforms and the ground truth (target) transforms. These cosine similarity curves show the same trend and support the same conclusions as with the transformation error curves reported earlier; this is not surprising as the transformation error and cosine similarity metrics are analytically related to each other as detailed in Section 4.3. For $N \geq 64^2 = N_i N_o$, the cosine similarity approaches 1, matching the target transformations using both the data-free ($K = 1$) and deep learning-based ($K = 4$) diffractive designs as shown in Figs. 1.d, S1.d, 4.d and S4.d.

To further shed light on the performance of these different diffractive designs, the estimated transformations and their differences (in phase and amplitude) from the target matrices ($\boldsymbol{A}$) are shown in Figs. 2, 5, and Supplementary Figs. S2 and S5 for different diffractive parameters. Similarly, examples of complex-valued input-output fields for different diffractive designs are compared in Figs. 3, 6 and Supplementary Figs. S3, S6 against the ground truth output fields (calculated using the target transformations), along with the resulting phase and amplitude errors at the output FOV. From these figures, it can be seen that both data-free ($K = 1$) and deep learning-based ($K = 4$) diffractive designs with the same total number of neurons can all-optically generate the desired transformation and output field patterns with negligible error when $N \geq 64^2 = N_i N_o$. For $N < N_i N_o$, on the other hand, the output field amplitude and phase profiles using deep learning-based diffractive designs show much better match to the ground truth output field profiles when compared to data-free, matrix pseudoinverse-based diffractive designs (see e.g., Figs. 3, 6).

In this subsection, we presented diffractive designs that successfully approximated arbitrary complex-valued transformations, where the individual elements of target $\boldsymbol{A}$ matrices (shown in Fig. 1.b, Supplementary Fig. S1.b, Fig. 4.b and Supplementary Fig. S4.b) were randomly and independently generated as described in Section 4.4. Our results confirm that, for a given total number of diffractive features/neurons $(N)$ available, building deeper diffractive networks where these neurons are distributed across multiple, successive layers, one following the other, can significantly improve the transformation error, output field accuracy and the diffraction efficiency of the whole system to all-optically implement an arbitrary, complex-valued target transformation between an input and output FOV. Starting with the following subsection, we focus on some task-specific all-optical transformations, which are frequently used in various optics and photonics applications.

### 2.4.2. 2D discrete Fourier transform

Here we show that the 2D Fourier transform operation can be performed using diffractive designs such that the complex field at output FOV reflects the 2D discrete Fourier transform of the input field. Compared to lens-based standard Fourier transform operations, diffractive surface-based solutions are not based on the paraxial approximation and offer a much more compact set-up (with a significantly smaller axial distance, e.g., $< 50\lambda$, between the input-output planes) and do not suffer from aberrations, which is especially important for larger input/output FOVs.

The associated transform matrix $(\boldsymbol{A})$ corresponding to 2D discrete Fourier transform, all-optical transformation error, cosine similarity of the resulting all-optical transforms with respect the ground truth, the output MSE and the diffraction efficiency are shown in Fig. 7. For all these curves and metrics, our earlier conclusions made in Section 2.4.1 are also applicable. Data-free $(K =$ $1)$ and deep learning-based $(K = 4)$ diffractive designs achieve accurate results at the output FOV for $N \geq N_i N_o = 64^2$, where the transformation error and the output MSE both approach to $\sim 0$ while the cosine similarity reaches $\sim 1$, as desired. In terms of the diffraction efficiency at the output FOV, similar to our earlier observations in the previous section, deep learning-based diffractive designs offer major advantages over data-free diffractive designs. Further advantages of deep learning-based diffractive designs over their data-free counterparts include significantly improved output MSE and reduced transformation error for $N < N_i N_o$, confirming our earlier conclusions made in Section 2.4.1.

To further show the success of the diffractive designs in approximating the 2D discrete Fourier transformation, in Fig. 8 we report the estimated transformations and their differences (in phase and amplitude) from the target 2D discrete Fourier transformation matrix for different diffractive designs. Furthermore, in Fig. 9, examples of complex-valued input-output fields for different diffractive designs are compared against the ground truth output fields (calculated using the 2D discrete Fourier transformation), along with the resulting phase and amplitude errors at the output FOV, all of which illustrate the success of the presented diffractive designs.

### 2.4.3. Permutation matrix-based transform

For a given randomly generated permutation matrix $(\boldsymbol{P})$, the task of the diffractive design is to all-optically obtain the permuted version of the input complex-field at the output FOV. Although the target ground truth matrix $(\boldsymbol{P})$ for this case is real-valued and relatively simpler compared to that of e.g., the 2D Fourier transform matrix, an all-optical permutation operation that preserves the phase and amplitude of each point is still rather unconventional and challenging to realize using standard optical components. To demonstrate this capability, we randomly selected a permutation matrix as shown in Fig. 10b and designed various diffractive surfaces to all-optically perform this target permutation operation at the output FOV. The performances of these data-free and data-driven, deep learning-based diffractive designs are compared in Figs. 10c-f. The success of the diffractive all-optical transformations, matching the target permutation operation is demonstrated when $N \geq N_i N_o$, revealing the same conclusions discussed earlier for the other transformation matrices that were tested. For example, deep learning-based diffractive designs ($K = 4$) with $N \geq N_i N_o$ neurons were successful in performing the randomly selected permutation operation all-optically, and achieved a transformation error and output MSE of $\sim 0$, together with a cosine similarity of $\sim 1$ (see Fig. 10). Estimated transforms and sample output patterns, together with their differences with respect to the corresponding ground truths are also reported in Figs. 11 and 12, respectively, further demonstrating the success of the presented diffractive designs.

### 2.4.4. High-pass filtered coherent imaging

In this sub-section, we present diffractive designs that perform high-pass filtered coherent imaging, as shown in Supplementary Fig. S7. This high-pass filtering transformation is based on the Laplacian operator described in Section 4.4. Similar to the 2D discrete Fourier transform operation demonstrated earlier, diffractive surface-based solutions to high-pass filtered coherent imaging are not based on a low numerical aperture assumption or the paraxial approximation, and provide an axially compact implementation with a significantly smaller distance between the input-output planes (e.g., $< 50\lambda$); furthermore, these diffractive designs can handle large input/output FOVs without suffering from aberrations.

Our results reported in Supplementary Fig. S7 also exhibit a similar performance to the previously discussed all-optical transformations, indicating that the pseudoinverse-based diffractive designs and the deep learning-based designs are successful in their all-optical approximation of the target transformation, reaching a transformation error and output MSE of $\sim 0$ for $N \geq N_i N_o$. Same as in other transformations that we explored, deep learning-based designs offer significant advantages compared to their data-free counterparts in the diffraction efficiency that is achieved at the output FOV. The estimated sample transformations and their differences from the ground truth transformation are shown in Supplementary Fig. S8.

Furthermore, as can be seen from the estimated output images and their differences with respect to the corresponding ground truth images (shown in Supplementary Fig. S9), the diffractive designs can accurately perform high-pass filtered coherent imaging for $N \geq N_i N_o$, and for $N < N_i N_o$ deep learning-based diffractive designs exhibit better accuracy in approximating the target output field, which are in agreement with our former observations in earlier sections.

## 3. Discussion

Through our results and analysis, we showed that it is possible to synthesize an arbitrary complex-valued linear transformation all-optically using diffractive surfaces. We covered a wide range of target transformations, starting from rather general cases, e.g. arbitrarily generated unitary, nonunitary (invertible) and noninvertible transforms, also extending to more specific transformations such as the 2D Fourier transform, 2D permutation operation as well as high-pass filtered coherent imaging operation. In all the linear transformations that we presented in this paper, the diffractive networks realized the desired transforms with negligible error when the total number of neurons reached $N \geq N_i N_o$. It is also important to note that the all-optical transformation accuracy of the deep learning-based diffractive designs improves as the number of diffractive layers is increased, e.g., from $K = 1, 2$ to $K = 4$. Despite sharing the same number of total neurons in each case (i.e., $N = \sum_{k=1}^{K} N_{L_k}$), deep learning-based diffractive designs prefer to distribute these $N$ trainable diffractive features/neurons into multiple layers, favoring deeper diffractive designs overall.

In addition to the all-optical transformation error, cosine similarity and output MSE metrics, the output diffraction efficiency is another very important metric as it determines the signal-to-noise ratio of the resulting all-optical transformation. When we compare the diffraction efficiencies of different networks, we observe that the data-free, matrix pseudoinverse-based designs perform the worst among all the configurations that we have explored (see Figs. 1.f, 4.f, 7.f, 10.f, and Supplementary Figs. S1.f, S4.f, S7.f). This is majorly caused by the larger magnitudes of the transmittance values of the neurons that are located close to the edges of the diffractive layer, when compared to the neurons at the center of the same layer. Since these "edge" neurons are further away from the input FOV, their larger transmission magnitudes ($|\boldsymbol{t}|$) compensate for the significantly weaker optical power that falls onto these edge neurons from the input FOV. Since we are considering here passive diffractive layers only, the magnitude of the transmittance value of an optical neuron cannot be larger than one (i.e., $|\boldsymbol{t}| \leq 1$), and therefore as the edge neurons in a data-free design start to get more transmissive to make up for the weaker input signals at their locations, the transmissivity of the central neurons of the diffractive layer become lower, balancing off their relative powers at the output FOV to be able to perform an arbitrary linear transformation. This is at heart of the poor diffraction efficiency that is observed with data-free, matrix pseudoinverse-based designs. In fact, the same understanding can also intuitively explain why deep learning-based diffractive designs prefer to distribute their trainable diffractive neurons into multiple layers. By dividing their total trainable neuron budget ($N$) into multiple layers, deeper diffractive designs (e.g., $K = 4$) avoid using neurons that are laterally further away from the center. This way, the synthesis of an arbitrary all-optical transformation can be achieved much more efficiently, without the need to weaken the transmissivity of the central neurons of a given layer. Stated differently, deep learning-based diffractive designs utilize a given neuron budget more effectively and can efficiently perform an arbitrary complex-valued transformation between an input and output FOV.

In fact, deep learning-based, data-driven diffractive designs can be made even more photon efficient by restricting each diffractive layer to be a phase-only element (i.e., $|\boldsymbol{t}| = 1$ for all the neurons) during the iterative learning process of a target complex-valued transformation. To demonstrate this capability with increased diffraction efficiency, we also designed diffractive networks with phase-only layers. The all-optical transformation performance metrics of these phase-only diffractive designs are summarized in Supplementary Figs. S10, S11, S12 and S13, corresponding to the same arbitrarily selected unitary transform (Fig. 1), nonunitary but invertible transform (Fig. 4), the 2D Fourier transform (Fig. 7) and the randomly selected permutation operation (Fig. 10), respectively. These results indicate that much better output diffraction efficiencies can be achieved using phase-only diffractive networks, with some trade-off in the all-optical transformation performance. The relative increase in the transformation errors and the output MSE values that we observed in phase-only diffractive networks is caused by the reduced degrees of freedom in the diffractive design since $|\boldsymbol{t}| = 1$ for all the neurons. Regardless, by increasing the total number of neurons $(N > N_i N_o)$, the phase-only diffractive designs approach the all-optical transformation performance of their complex-valued counterparts designed by deep learning, while also providing a much better diffraction efficiency at the output FOV (see Supplementary Figs. S10, S11, S12 and S13). Note also that, while the phase-only diffractive layers are individually lossless, the forward propagating optical fields still experience some power loss due to the opaque regions that are assumed to surround the diffractive surfaces (which is a design constraint as detailed in Section 2.1). In addition to these losses, the field energy that lies outside of the output FOV is also considered a loss from the perspective of the target transformation, which is defined between the input and output FOVs.

In order to further increase the output diffraction efficiency of our designs, we adopted an additional strategy that includes a diffraction efficiency-related penalty term in the loss function (detailed in Section 4.5). Performance quantification of these diffraction efficient designs that have different loss function parameters is shown in Supplementary Fig. S14 for a unitary transform; the same figure also includes the comparison between the complex-valued and phase-only modulation schemes that do not have a diffraction efficiency-related penalty term in their loss functions. The analyses shown in Supplementary Fig. S14 indicate that the output diffraction efficiency of the network can be significantly improved and maintained almost constant among different designs with different numbers of neurons by the introduction of the diffraction efficiency-related penalty term during the training. Moreover, the transformation error, cosine similarity and MSE performance of the complex-valued modulation scheme that has no diffraction efficiency term can be approximately matched through this new training strategy that includes an additional loss term dedicated to improve diffraction efficiency.

We also compared our results to a data-based iterative projection method, which can be considered as an extension of the Gerchberg-Saxton algorithm for multi-layer settings[45]. Using this iterative projection method,[45] we trained phase-only transmittance values of the diffractive layers in order to approximate the first unitary transform presented in Fig. 1 by feeding them with the corresponding ground truth input/output fields. Then, we computed the associated transformation error and the other performance metrics through the projection-based optimized neuron values. We report the results of this comparison in Supplementary Fig. S15. The transformation error, cosine similarity measure and MSE results indicate that as the number of diffractive layers increases, the deep learning-based all-optical solutions become much more accurate in their approximation compared to the performance of iterative projection-based diffractive designs.

We also report the convergence analysis of our deep learning-based designs in Supplementary Fig. S16. The simulations for this convergence analysis were carried out for $K = 4$ using both the complex-valued and phase-only modulation layers when $N = 48^2$ , $64^2$ , or $80^2$ . The MSE curves in Supplementary Fig. S16 show a monotonic decrease as the number of epochs increases during the training. On the other hand, although the general trends of the transformation error and the cosine similarity metrics are almost monotonic, they show some local variations between consecutive training epochs, as shown in Supplementary Figs. S16a-b. This behavior is expected since the diffractive network does not directly minimize the transformation error during its training; in fact, it minimizes the output MSE as given by Equation 10. In addition to these, we observe that the diffractive network design converges after a few epochs, where the training in the subsequent epochs serves for the purpose of fine tuning of the network's approximation performance.

In our analysis reported so far, there are some practical factors that are not taken into account as part of our forward optical model, which might degrade the performance of diffractive networks: (1) material absorption, (2) surface reflections and (3) fabrication imperfections. By using materials with low loss and appropriately selected 3D fabrication methods, these effects can be made negligible compared with the optical power of the forward propagating modes within the diffractive network. Alternatively, one can also include such absorption- and reflection-related effects as well as mechanical misalignments (or fabrication imperfections) as part of the forward model of the optical system, which can be better taken into account during the deep learning-based optimization of the diffractive layers. Importantly, previous experimental studies[28,29,32,34] reported on various diffractive network applications indicate that the impact of fabrication errors, reflection and absorption-based losses are indeed small and do not create a significant discrepancy between the predictions of the numerical forward models and the corresponding experimental measurements.

Finally, we should emphasize that for diffractive networks that have more than one layer, the transmittance values of the neurons of different layers appear in a coupled, multiplicative nature within the corresponding matrix-vector formulation of the all-optical transformation between the input and output FOVs [35]. Hence, a one-step, matrix pseudoinverse-based design strategy cannot be applied for multi-layered diffractive networks in finding all the neuron transmittance values. Moreover, for diffractive designs with a large $N$, the sizes of the matrices that need to undergo the pseudoinverse operation grow exponentially, which drastically increases the computational load and may prevent performing matrix pseudoinverse computations due to limited computer memory and computation time. This also emphasizes another important advantage of the deep learning-based design methods which can handle much larger number of diffractive neurons to be optimized for a given target transformation, thanks to the efficient error back-propagation algorithms and computational tools that are available. Similarly, if $N_i$ and $N_o$ are increased as the sizes of the input and output FOVs are enlarged, the total number of diffractive neurons needed to successfully approximate a given complex-valued linear transformation will accordingly increase to $D = N_i N_o$, which indicates the critical number of total diffractive features (marked with the vertical dashed lines in our performance metrics related figures).

# 4. Materials and methods

## 4.1. Diffractive network configurations

In our numerical simulations, the chosen input and output FOV sizes are both $8 \times 8$ pixels. Hence, the target linear transforms, i.e., $\boldsymbol{A}$ matrices, have a size of $N_o \times N_i = 64 \times 64$. For a diffractive design that has a single diffractive surface ($K = 1$), the chosen axial distances are $d_1 = \lambda$ and $d_2 = 4\lambda$. For the networks that have two diffractive surfaces ($K = 2$), the chosen axial distances are $d_1 = \lambda$ and $d_2 = d_3 = 4\lambda$. Finally, for the 4-layered diffractive network ($K = 4$), the axial distances are chosen as $d_1 = \lambda$ and $d_2 = d_3 = d_4 = 4\lambda$. These axial distances can be arbitrarily changed without changing the conclusions of our analysis; they were chosen large enough to neglect the near-field interactions between successive diffractive layers, and small enough to perform optical simulations with a computationally feasible wave propagation window size. We chose our 2D wave propagation window as $N_d \times N_d = 144 \times 144$, which ends up with a size of $144^2 \times 144^2$ for $\boldsymbol{H}_d$ matrices, resulting in $\sim 430$ Million entries in each $\boldsymbol{H}_d$ matrix.

## 4.2. Image datasets and diffractive network training parameters

To obtain the diffractive surface patterns that collectively approximate the target transformation using deep learning-based training, we generated a complex-valued input-output image dataset for each target $\boldsymbol{A}$. To cover a wide range of spatial patterns, each input image in the dataset has a different sparsity ratio with randomly chosen pixel values. We also included rotated versions of each training image. We can summarize our input image dataset as

$$(4P + 8P + 16P + 32P + 48P + 64P) \times 4R \times S \quad 15$$

where $S$ refers to the number of images for each sub-category of the training image set defined by $kP$ for $k \in \{4,8,16,32,48,64\}$, which indicates a training image where $k$ pixels out of $N_i$ pixels are chosen to be nonzero (with all the rest of the pixels being zero). Hence, $k$ indicates the fill factor for a given image, as shown in Supplementary Fig. S17. We choose $S = 15{,}000$ for the training and $S = 7{,}500$ for the test image sets. Also, $4R$ in Equation 15 indicates the four different image rotations of a given training image, where the rotation angles are determined as $0°, 90°, 180°$ and $270°$. For example, $16P$ in Equation 15 indicates that randomly chosen 16 pixels out of 64 pixels of an image are nonzero and the remaining 48 pixels are zero. Following this formalism, we generated a total of $360K$ images for the training dataset and $180K$ for the test image dataset. Moreover, if a pixel in an image was chosen as nonzero, it took an independent random value from the set $\left\{\frac{32}{255}, \frac{33}{255}, \cdots, \frac{254}{255}, \frac{255}{255}\right\}$. Here the lower bound was chosen so that the "on" pixels can be well-separated from the zero-valued "off" pixels.

In this paper, we used the same input ($\boldsymbol{i}$) image dataset for all the transformation matrices ($\boldsymbol{A}$) that we utilized as our testbed. However, since the chosen linear transforms are different, the ground truth output fields are also different in each case, and were calculated based on $\boldsymbol{o} = \boldsymbol{A}\boldsymbol{i}$. Sample images from the input fields can be seen in Supplementary Fig. S17.

As discussed in Section 2.3, our forward model implements Equation 2 and the DFT operations are performed using the fast Fourier transform algorithm[42]. In our deep learning models, we chose the loss function as shown in Equation 10. All the networks were trained using Python (v3.6.5) and TensorFlow (v1.15.0, Google Inc.), where the Adam optimizer was selected during the training. The learning rate, batch size and the number of training epochs were set to be 0.01, 8 and 50, respectively.

## 4.3. Computation of all-optical transformation performance metrics

As our all-optical transformation performance metrics, we used (i) the transformation error, (ii) the cosine similarity between the ground truth and the estimate transformation matrices, (iii) normalized output MSE and (iv) the mean output diffraction efficiency.

The first metric, the transformation error, is defined in Equation 5, which was used for both pseudoinverse-based diffractive designs and deep learning-based designs. The second chosen metric is the cosine similarity between two complex-valued matrices, which is defined as

$$U(\boldsymbol{a}, \widehat{\boldsymbol{a}}') = \frac{|\langle \boldsymbol{a}, \widehat{\boldsymbol{a}}' \rangle|}{\sqrt{\sum_{l=1}^{N_i N_o} |\boldsymbol{a}[l]|^2} \sqrt{\sum_{l=1}^{N_i N_o} |\widehat{\boldsymbol{a}}'[l]|^2}} = \frac{|\boldsymbol{a}^H \widehat{\boldsymbol{a}}'|}{\sqrt{\|\boldsymbol{a}\|^2} \sqrt{\|\widehat{\boldsymbol{a}}'\|^2}} \qquad 16$$

We use the notation $U(\boldsymbol{A}, \widehat{\boldsymbol{A}}')$ interchangeably with $U(\boldsymbol{a}, \widehat{\boldsymbol{a}}')$, both referring to Equation 16. Note that, even though the transformation error and cosine similarity metrics that are given by Equations 5 and 16, respectively, are related to each other, they end up with different quantities. The relationship between these two metrics can be revealed by rewriting Equation 5 as

$$\|\boldsymbol{a} - \widehat{\boldsymbol{a}}'\|^2 = (\boldsymbol{a} - \widehat{\boldsymbol{a}}')^H (\boldsymbol{a} - \widehat{\boldsymbol{a}}') = \|\boldsymbol{a}\|^2 + \|\widehat{\boldsymbol{a}}'\|^2 - 2Re\{\boldsymbol{a}^H \widehat{\boldsymbol{a}}'\} \qquad 17$$

where $Re\{\cdot\}$ operator extracts the real part of its input. As a result, apart from the vector normalization constants, $\|\boldsymbol{a}\|^2$ and $\|\widehat{\boldsymbol{a}}'\|^2$, Equations 16 and 17 deal with the magnitude and real part of the inner product $(\boldsymbol{a}^H \widehat{\boldsymbol{a}}')$, respectively.

For the third metric, the normalized MSE calculated at the output FOV, we used the following equation:

$$\begin{aligned} E[\|\widetilde{\boldsymbol{o}} - \widetilde{\boldsymbol{o}}'\|^2] &= \frac{1}{S_T N_o} \sum_{s=1}^{S_T} \sum_{l=1}^{N_o} |\widetilde{\boldsymbol{o}}_s[l] - \widetilde{\boldsymbol{o}}'_s[l]|^2 \\ &= \frac{1}{S_T N_o} \sum_{s=1}^{S_T} \sum_{l=1}^{N_o} |\sigma_s \boldsymbol{o}_s[l] - \sigma'_s \boldsymbol{o}'_s[l]|^2 \end{aligned} \qquad 18$$

where $E[\cdot]$ is the expectation operator and $S_T$ is the total number of the image samples in the test dataset. The vectors $\boldsymbol{o}_s = \boldsymbol{A}\boldsymbol{i}_s$ and $\boldsymbol{o}'_s = \boldsymbol{A}'\boldsymbol{i}_s$ represent the ground truth and the estimated output fields (at the output FOV), respectively, for the $s^{th}$ input image sample in the dataset, $\boldsymbol{i}_s$. The normalization constant, $\sigma_s$ is given by Equation 12, and $\sigma_s'$, can be computed from Equation 13, by replacing $\sigma_s'$ and $\boldsymbol{o}_s'$ by $\sigma_{s,c}'$ and $\boldsymbol{o}_{s,c}'$, respectively.

Finally, we chose the mean diffraction efficiency of the diffractive system as our last performance metric, which is computed as

$$E\left[\frac{\|\boldsymbol{o}'\|^2}{\|\boldsymbol{i}\|^2}\right] = \frac{1}{S_T}\sum_{s=1}^{S_T}\frac{\sum_{l=1}^{N_o}|\boldsymbol{o}_s'[l]|^2}{\sum_{l=1}^{N_i}|\boldsymbol{i}_s[l]|^2} \tag{19}$$

## 4.4. Generation of ground truth transformation matrices

To create the unitary transformations, as presented in Fig. 1.b and Supplementary Fig. S1.b, we first generated a complex-valued Givens rotation matrix, which is defined for a predetermined $i, j \in \{1,2, \cdots, N_i\}$ and $i \neq j$ pair as

$$\boldsymbol{R}_{ij}[n,m] = \begin{cases} 1, & \text{if } n = m, n \neq i \text{ and } n \neq j \\ e^{j\theta_1}\cos\theta_3, & \text{if } n = m = i \\ e^{-j\theta_1}\cos\theta_3, & \text{if } n = m = j \\ e^{j\theta_2}\sin\theta_3, & \text{if } n = i \text{ and } m = j \\ -e^{-j\theta_2}\sin\theta_3, & \text{if } n = j \text{ and } m = i \\ 0, & \text{otherwise} \end{cases} \tag{20}$$

where $\theta_1, \theta_2, \theta_3 \in [0,2\pi)$ are *randomly* generated phase values. Then a unitary matrix was computed as

$$\boldsymbol{R} = \prod_{t=1}^{T}\boldsymbol{R}_{i_t j_t} \tag{21}$$

where $(i_t, j_t)$ pair is randomly chosen for each $t$. We used $T = 10^5$ in our simulations. As a result, for each $t$ in Equation 21, $(i_t, j_t)$ and $(\theta_1, \theta_2, \theta_3)$ were chosen randomly. It is straightforward to show that the resulting $\boldsymbol{R}$ matrix in Equation 21 is a unitary matrix.

To compute the nonunitary but invertible matrices, we first generated two unitary matrices $\boldsymbol{R}_U$ and $\boldsymbol{R}_V$, as described by Equations 20 and 21, and then a diagonal matrix $\boldsymbol{X}$. The diagonal elements of $\boldsymbol{X}$ takes uniformly, independently and identically generated random real values in the range $[0.3,1]$, where the lower limit is determined to be large enough to prevent numerical instabilities and the upper limit is determined to prevent amplification of the orthonormal components of $\boldsymbol{R}_U$ and $\boldsymbol{R}_V$. Then, the nonunitary but invertible matrix is generated as $\boldsymbol{R}_U\boldsymbol{X}\boldsymbol{R}_V^H$, which is in the form of the SVD of the resulting matrix. It is straightforward to show that the resulting matrix is invertible. However, to make sure that it is nonunitary, we numerically compute its Hermitian and its inverse separately, and confirm that they are not equal. Similarly, to compute the noninvertible transformation matrix, as shown in e.g., Supplementary Fig. S4.b, we equated the randomly chosen half of the diagonal elements of $\boldsymbol{X}$ to zero and randomly chose the remaining half to be in the interval $[0.3,1]$. Following this, we computed the noninvertible matrix as $\boldsymbol{R}_U\boldsymbol{X}\boldsymbol{R}_V^H$, by re-computing new unitary matrices $\boldsymbol{R}_U$ and $\boldsymbol{R}_V$, which end up to be completely different from the $\boldsymbol{R}_U$ and $\boldsymbol{R}_V$ matrices that were computed for the nonunitary and invertible transform.

The 2D DFT operation for the square input aperture located at the center of the input plane was defined by

$$\boldsymbol{o}_{\boldsymbol{2D}}[p,q] = \frac{1}{\sqrt{N_i}} \sum_{n=-\frac{\sqrt{N_i}}{2}}^{\frac{\sqrt{N_i}}{2}-1} \sum_{m=-\frac{\sqrt{N_i}}{2}}^{\frac{\sqrt{N_i}}{2}-1} \boldsymbol{i}_{\boldsymbol{2D}}[n,m] e^{-j\frac{2\pi}{\sqrt{N_i}}(pn+qm)} \qquad (22)$$

where $\boldsymbol{i}_{\boldsymbol{2D}}$ and $\boldsymbol{o}_{\boldsymbol{2D}}$ represent the 2D fields on the input and output FOVs, respectively, and $n,m,p,q \in \left\{-\frac{\sqrt{N_i}}{2}, -\frac{\sqrt{N_i}}{2}+1, \cdots, \frac{\sqrt{N_i}}{2}-1\right\}$. Here we assume that the square-shaped input and output FOVs have the same area and number of pixels, i.e., $N_i = N_o$. Moreover, since we assume that the input and output FOVs are located at the center of their associated planes, the space and frequency indices start from $-\sqrt{N_i}/2$. Therefore, the $\boldsymbol{A}$ matrix associated with the 2D centered DFT, which is shown in Fig. 7.b, performs the transform given in Equation 22.

The permutation ($\boldsymbol{P}$) operation performs a one-to-one mapping of the complex-value of each pixel on the input FOV onto a different location on the output FOV. Hence the randomly selected transformation matrix ($\boldsymbol{A} = \boldsymbol{P}$) associated with the permutation operation has only one nonzero element along each row, whose value equals to 1, as shown in Fig. 10.b.

Finally, the transformation matrix corresponding to the high-pass filtering operation, as shown in Supplementary Fig. S7.b, is generated from the Laplacian high-pass filter whose 2D convolution kernel is

$$\begin{bmatrix} 1 & 4 & 1 \\ 4 & -20 & 4 \\ 1 & 4 & 1 \end{bmatrix} \qquad 23$$

After generating the 2D matrix by applying the appropriate vectorization operation, we also normalize the resulting matrix with its largest singular value, to prevent the amplification of the orthonormal components.

### 4.5. Penalty Term for Improved Diffraction Efficiency

To increase the diffraction efficiency at the output FOV of a diffractive network design, we used the following modified loss function:

$$L = c_M L_M + c_D L_D \qquad 24$$

where $L_M$ is the MSE loss term which is given in Equation 10 and $L_D$ is the additional loss term that penalizes poor diffraction efficiency:

$$L_D = e^{-\alpha X} \qquad 25$$

where $X$ is the diffraction efficiency term which is given by Equation 19. In Equations 24 and 25, $c_M$, $c_D$ and $\alpha$ are the user-defined weights. In earlier designs where the diffraction efficiency has not been penalized or taken into account during the training phase, $c_M$ and $c_D$ were taken as 1 and 0, respectively.

**Acknowledgements:** The authors acknowledge the US Air Force Office of Scientific Research (AFOSR), Materials with Extreme Properties Program funding (FA9550-21-1-0324). O.K. acknowledges the partial support of the Fulbright Commission of Turkey.

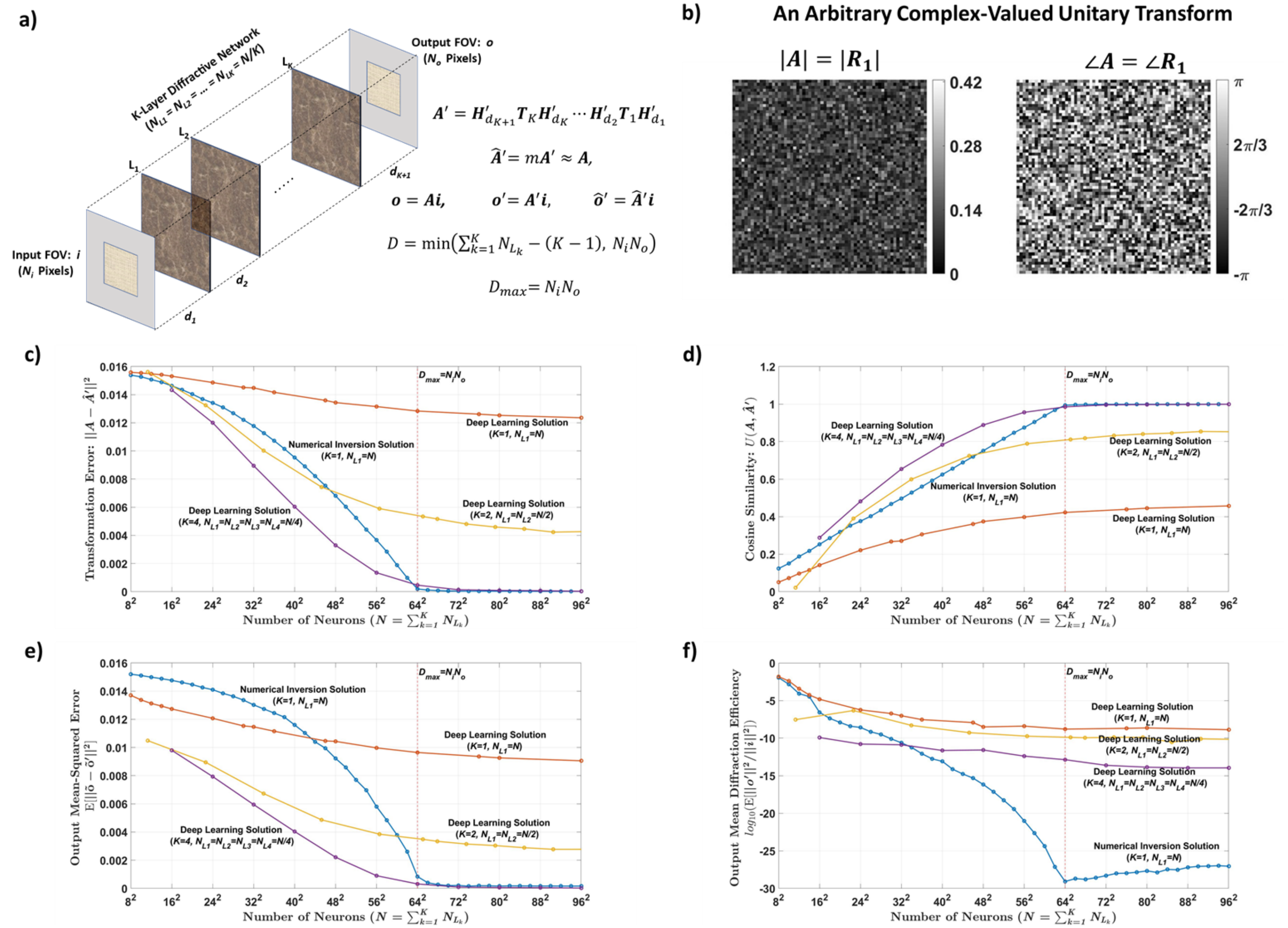

***Figure 1: Diffractive all-optical transformation results for an arbitrary complex-valued unitary transform. a.*** *Schematic of a K-layer diffractive network, that all-optically performs a linear transformation between the input and output fields-of-views that have* $N_i$ *and* $N_o$ *pixels, respectively. The all-optical transformation matrix due to the diffractive layer(s) is given by* $\boldsymbol{A'}$*.* ***b.*** *The magnitude and phase of the ground truth (target) input-output transformation matrix, which is an arbitrarily generated complex-valued unitary transform, i.e.,* $\boldsymbol{R_1^H R_1 = R_1 R_1^H = I}$*.* ***c.*** *All-optical transformation errors (see Equation 5). The x-axis of the figure shows the total number of neurons* $(N)$ *in a K-layered diffractive network, where each diffractive layer includes* $N/K$ *neurons. Therefore, for each point on the x-axis, the comparison among different diffractive designs (colored curves) is fair as each diffractive design has the same total number of neurons available. The simulation data points are shown with dots and the space between the dots are linearly interpolated.* ***d.*** *Cosine similarity between the vectorized form of the target transformation matrix in (b) and the resulting all-optical transforms (see Equation 16).* ***e.*** *Output MSE between the ground-truth output fields and the estimated output fields by the diffractive network (see Equation 18).* ***f.*** *The diffraction efficiency of the designed diffractive networks (see Equation 19).*

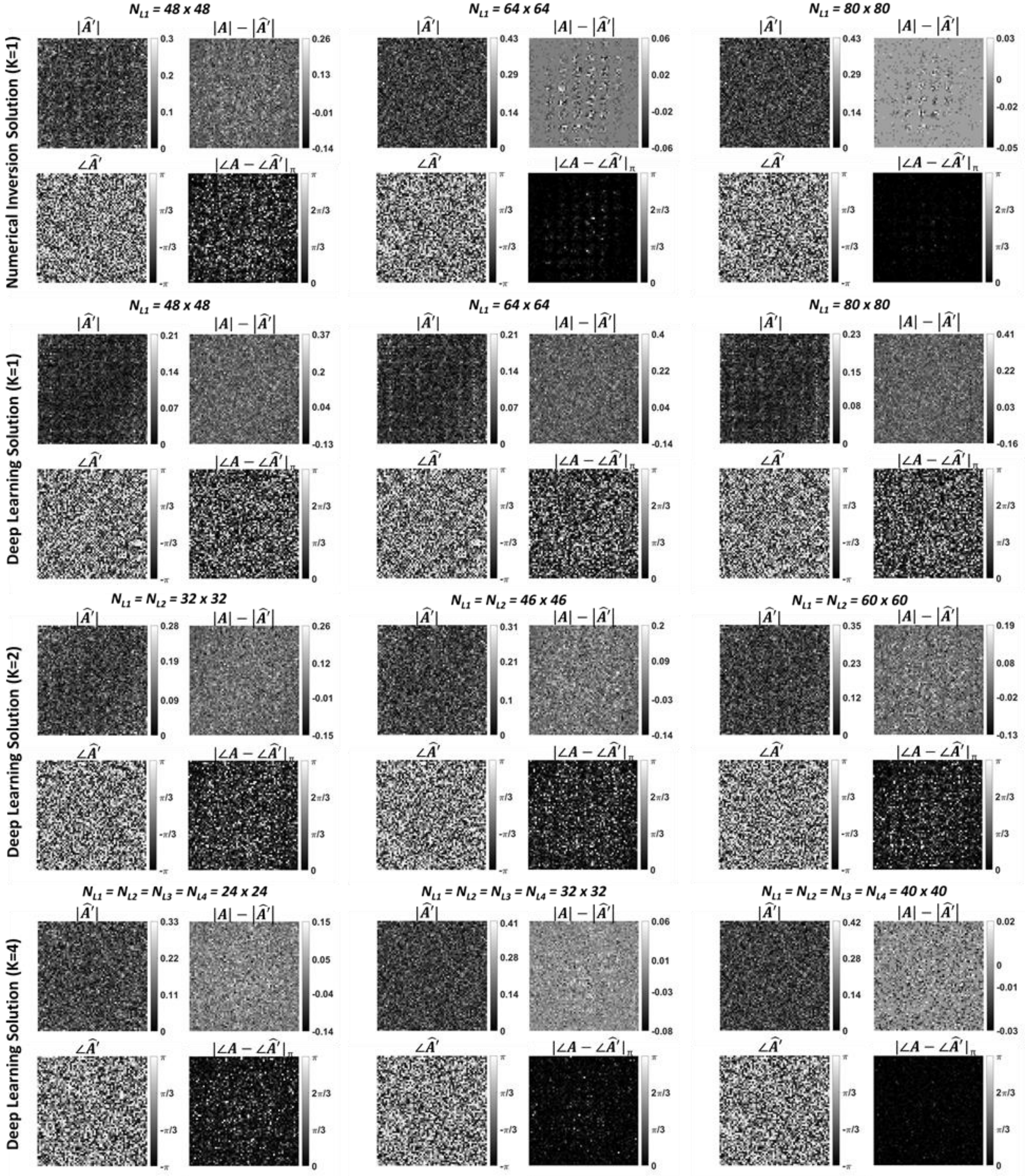

***Figure 2. Diffractive all-optical transformations and their differences from the ground truth, target transformation $(A)$ presented in Fig. 1.b.*** $\left|\angle A - \angle\widehat{A}'\right|_\pi$ *indicates the wrapped phase difference between the ground truth and the normalized all-optical transformation.*

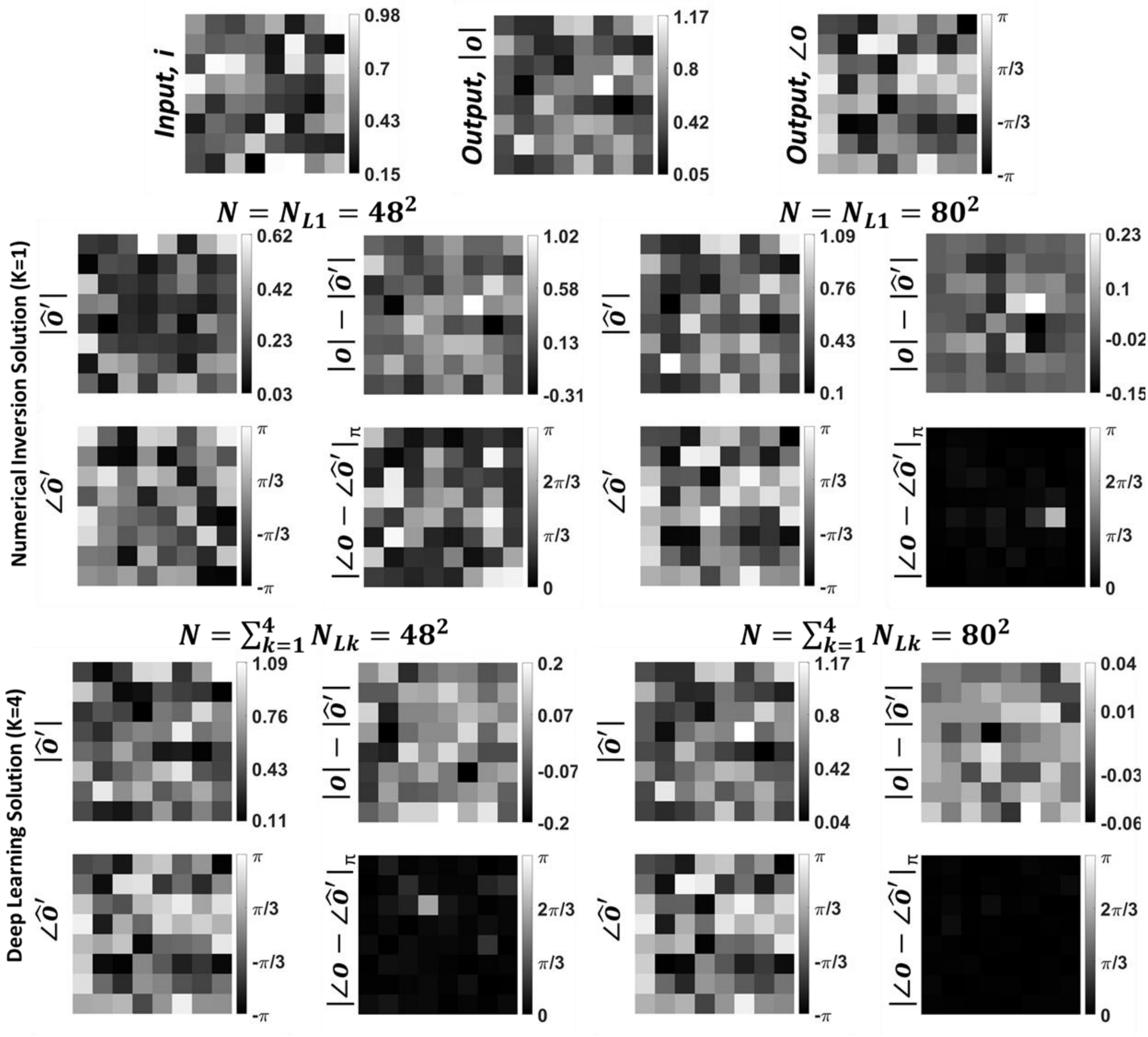

***Figure 3. Sample input-output images for the ground truth transformation presented in Fig. 1.b and the optical outputs by the diffractive designs for two different choices of $N$ ($N = 48^2$ and $N = 80^2$).*** *The magnitude and phase of the normalized output fields and the differences of these quantities with respect to the ground truth are shown.* $|\angle \boldsymbol{o} - \angle \widehat{\boldsymbol{o}}'|_\pi$ *indicates wrapped phase difference between the ground truth and the normalized output field.*

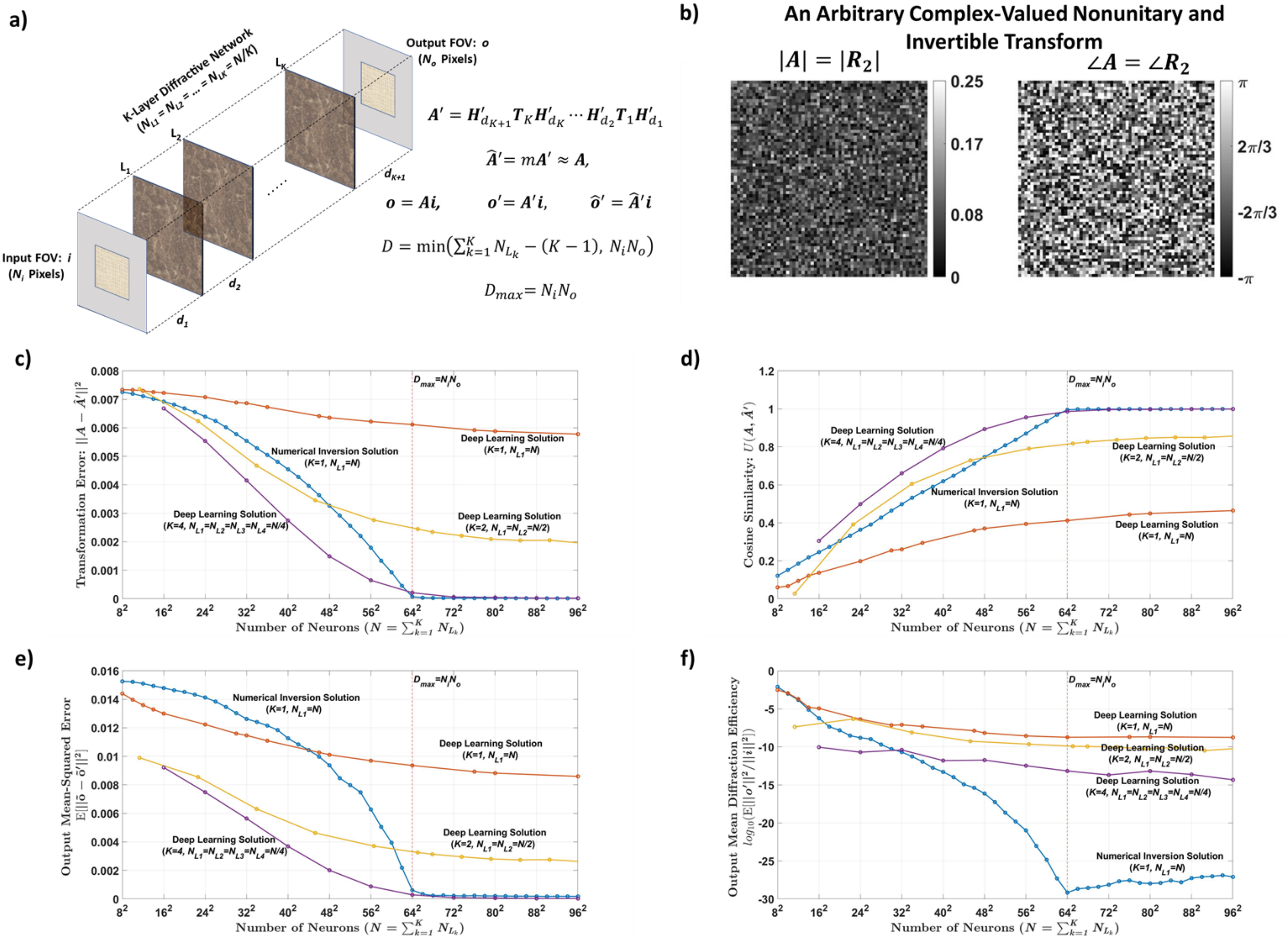

***Figure 4:*** *Follows the caption of Fig. 1, except that the target* $(\boldsymbol{A} = \boldsymbol{R_2})$ *is an arbitrary complex-valued nonunitary and invertible transform.*

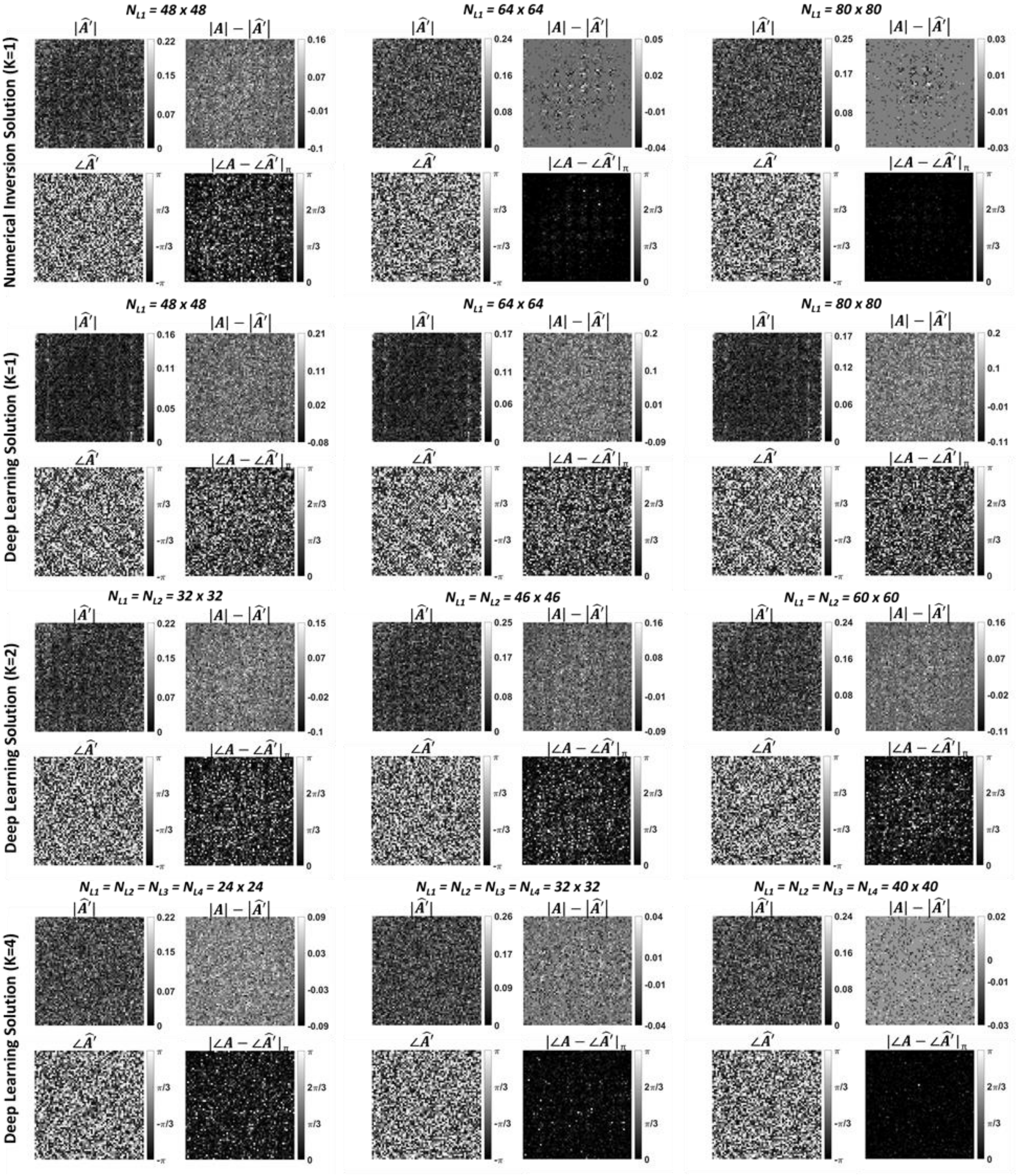

***Figure 5:*** *Follows the caption of Fig. 2, except that the target* $(\boldsymbol{A} = \boldsymbol{R_2})$ *is an arbitrary complex-valued nonunitary and invertible transform.*

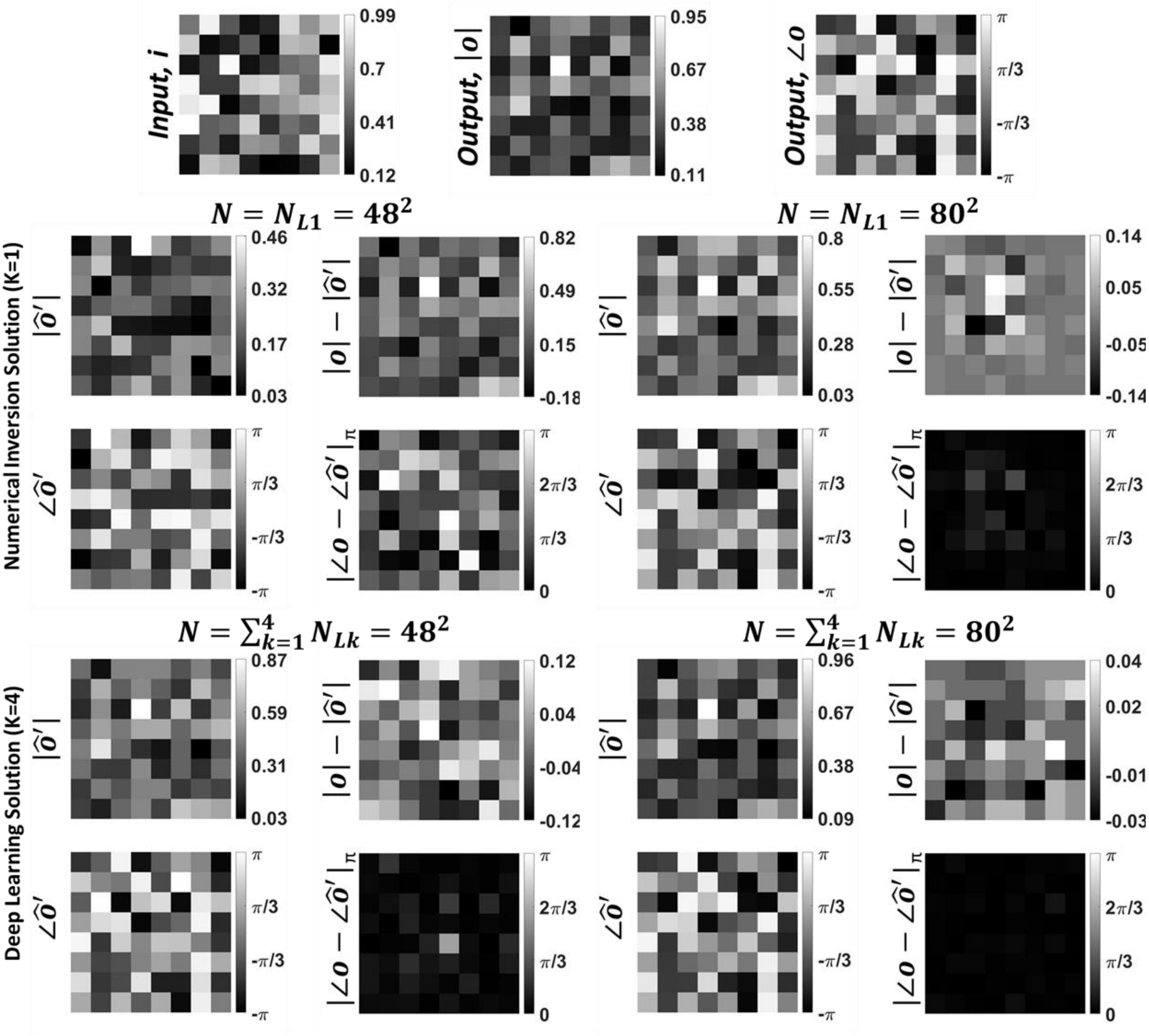

***Figure 6*** *Follows the caption of Fig. 3, except that the target ($A = R_2$) is an arbitrary complex-valued nonunitary and invertible transform.*

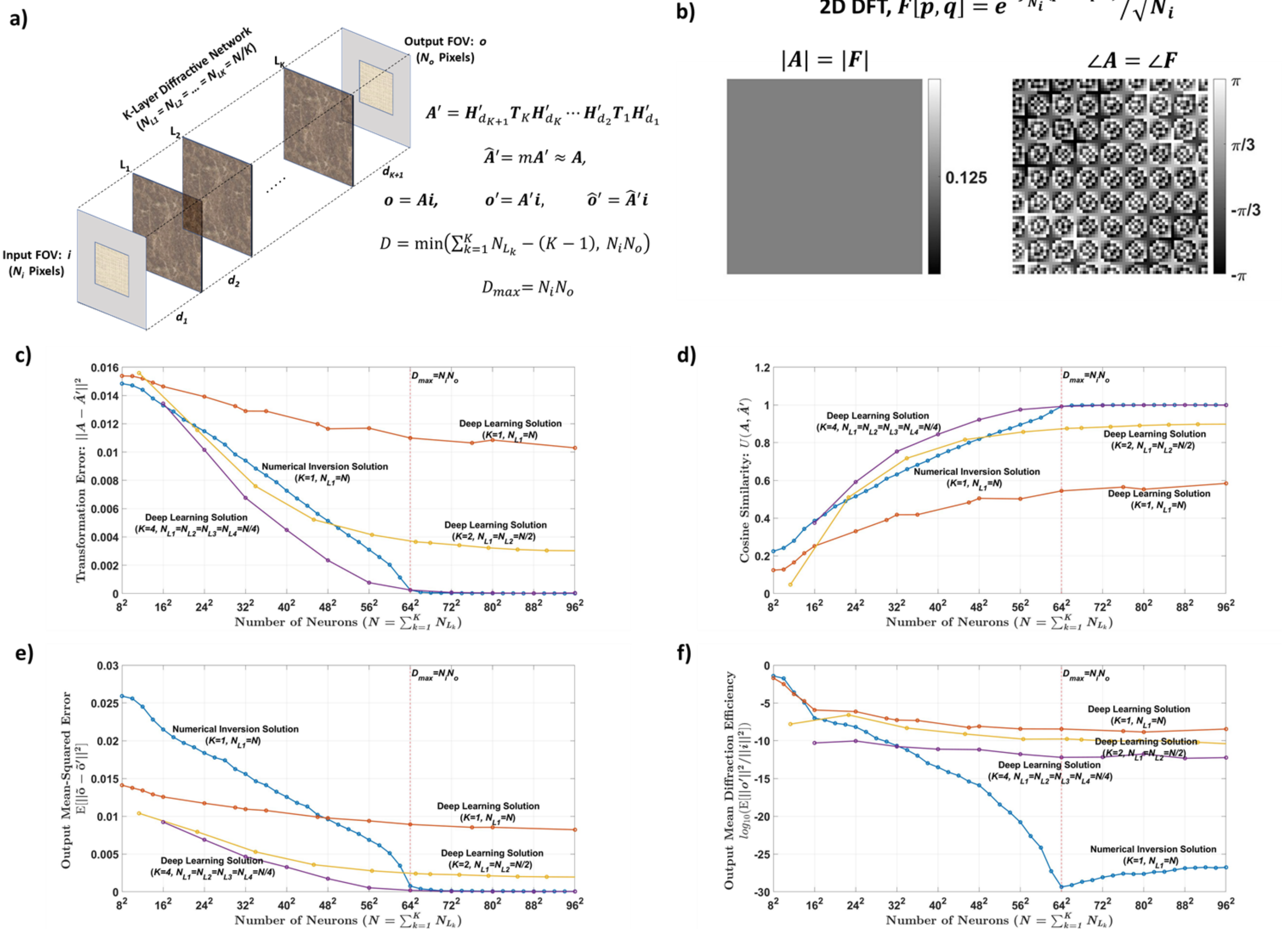

*Figure 7:* *Follows the caption of Fig. 1, except that the target ($A = F$) is the 2D discrete Fourier transform.*

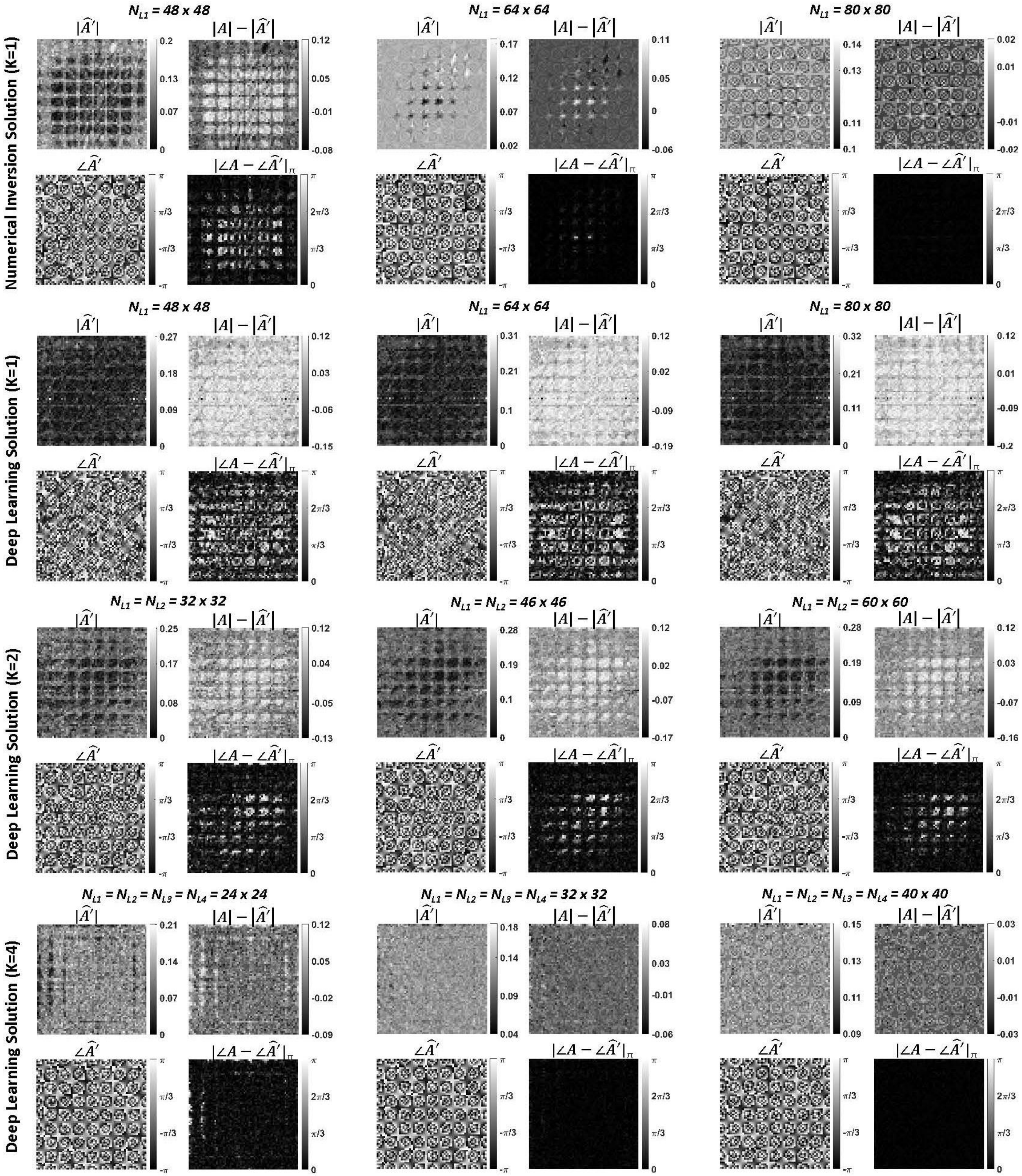

***Figure 8:*** *Follows the caption of Fig. 2, except that the target* $(\boldsymbol{A} = \boldsymbol{F})$ *is the 2D discrete Fourier transform.*

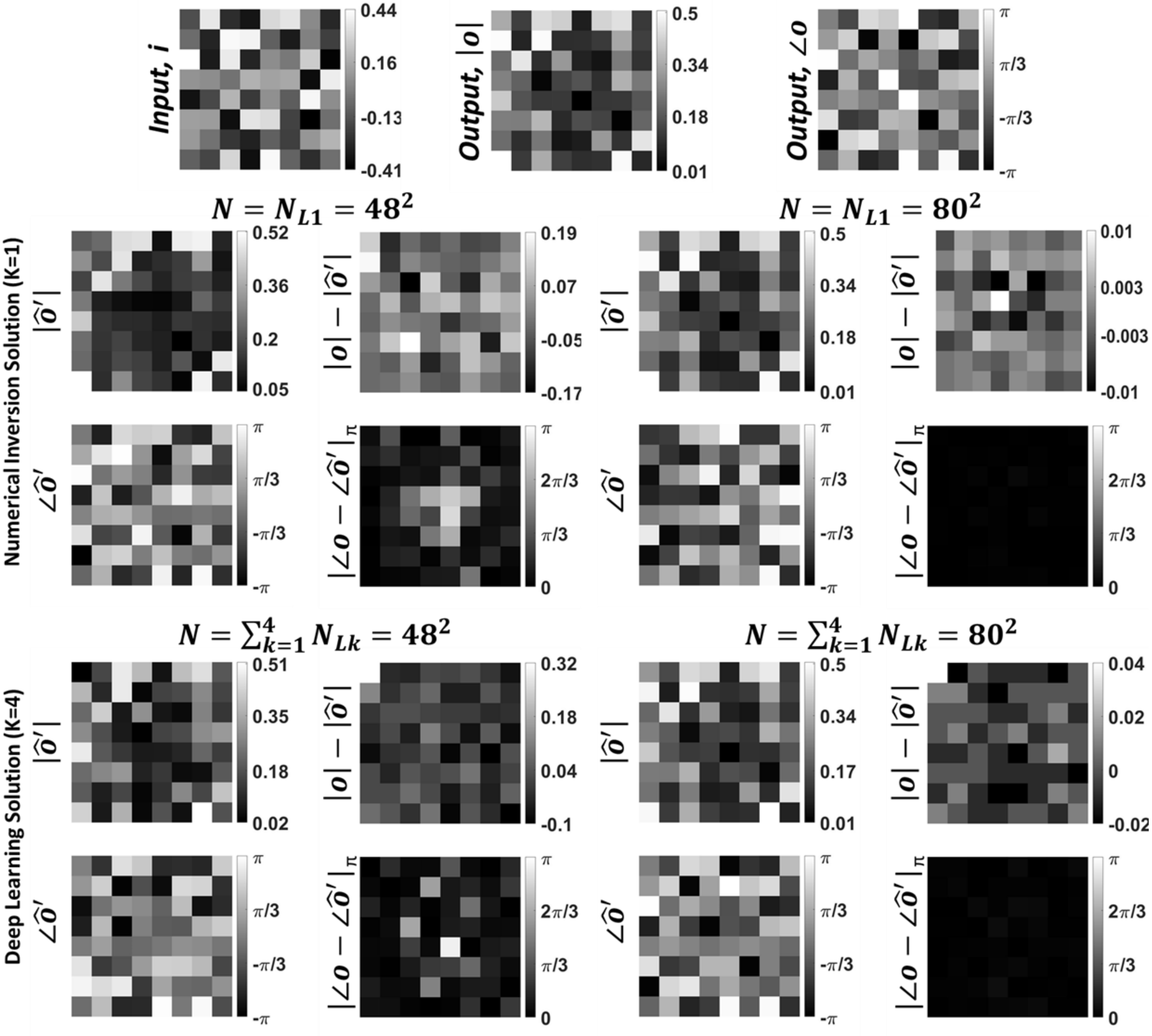

***Figure 9.*** *Follows the caption of Fig. 3, except that the target* $(\boldsymbol{A} = \boldsymbol{F})$ *is the 2D discrete Fourier transform.*

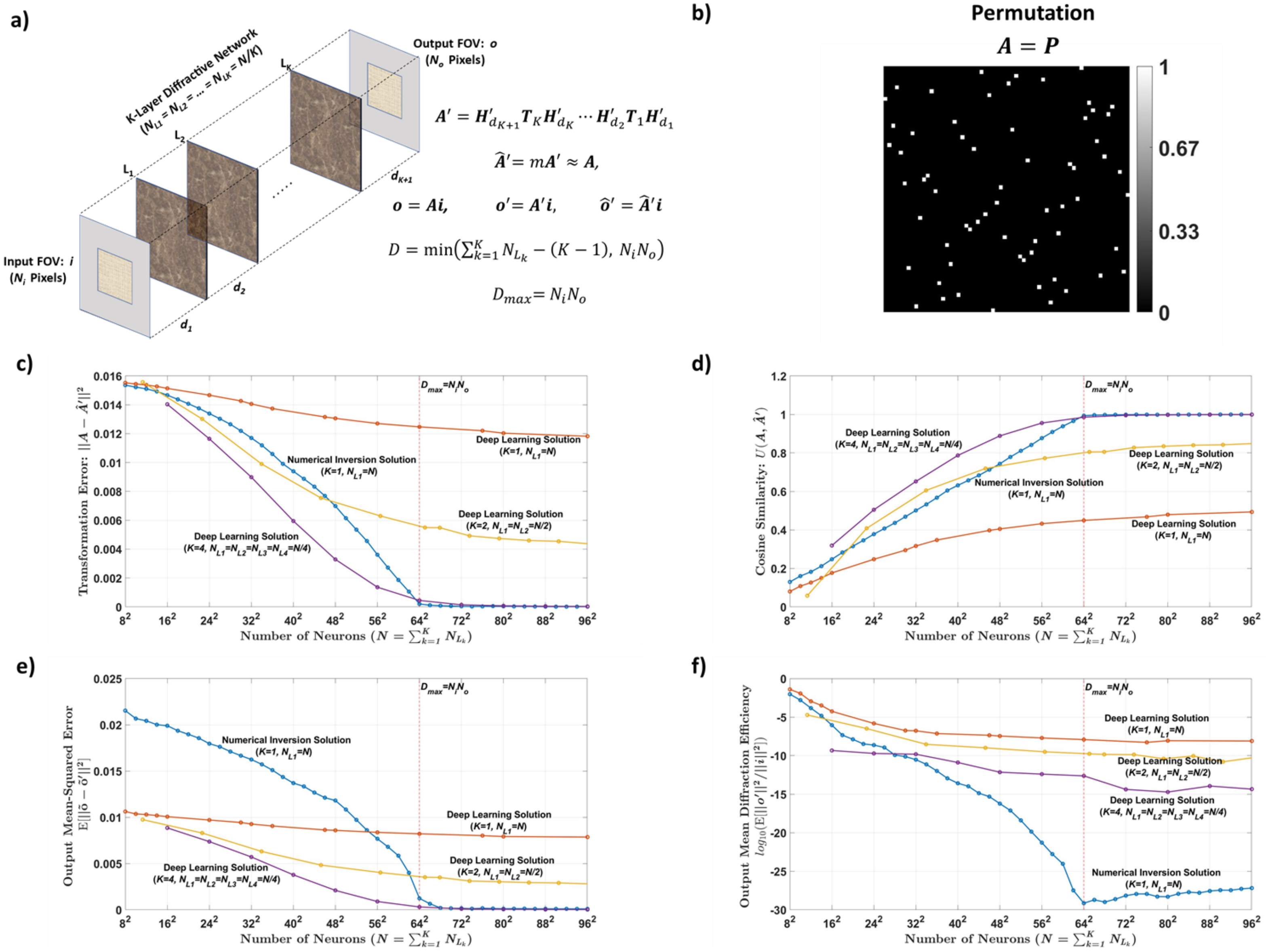

***Figure 10:*** *Follows the caption of Fig. 1, except that the target* $(\mathbf{A} = \boldsymbol{P})$ *is a permutation matrix.*

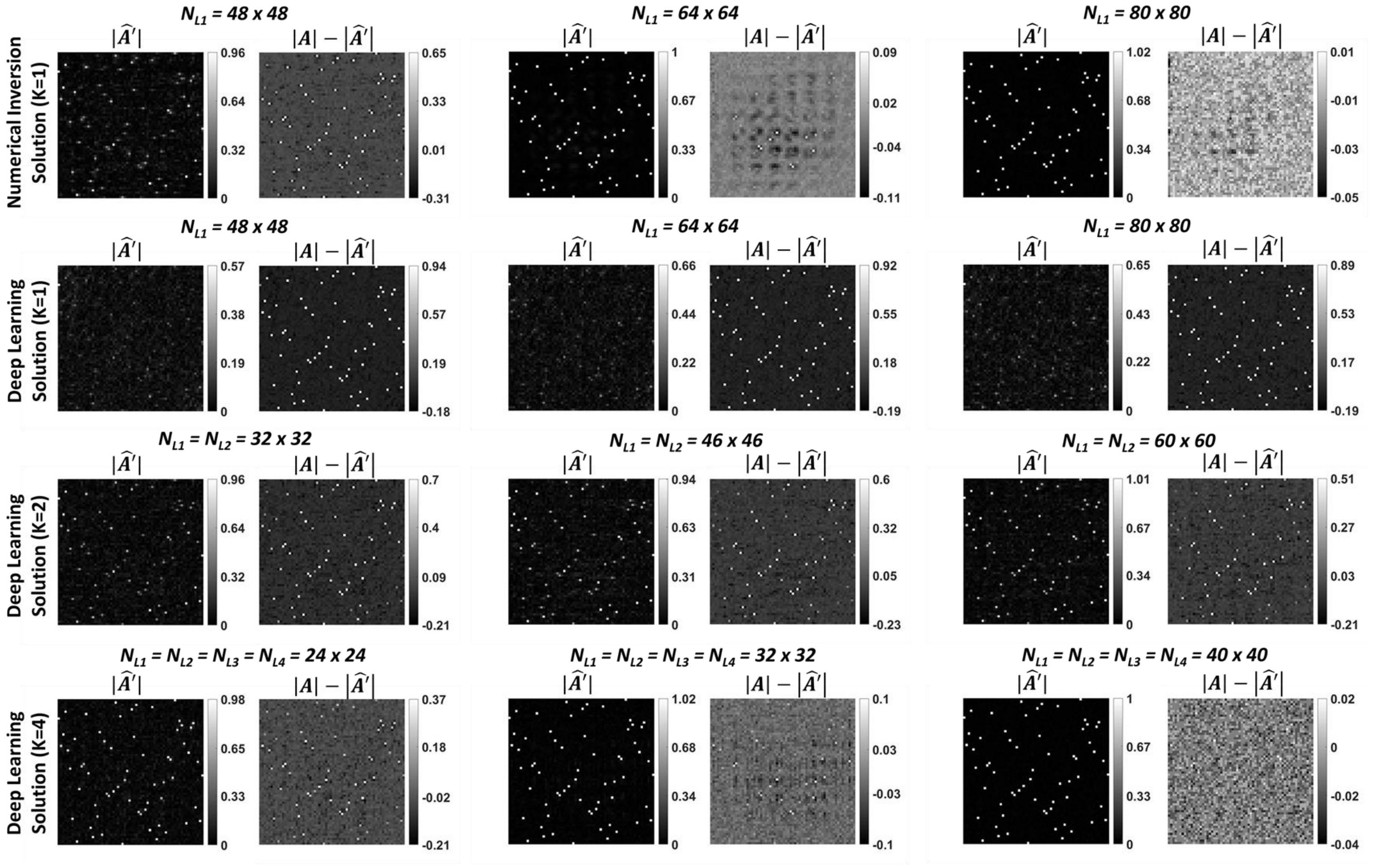

***Figure 11:*** *Follows the caption of Fig. 2, except that the target* $(\boldsymbol{A} = \boldsymbol{P})$ *is a permutation matrix.*

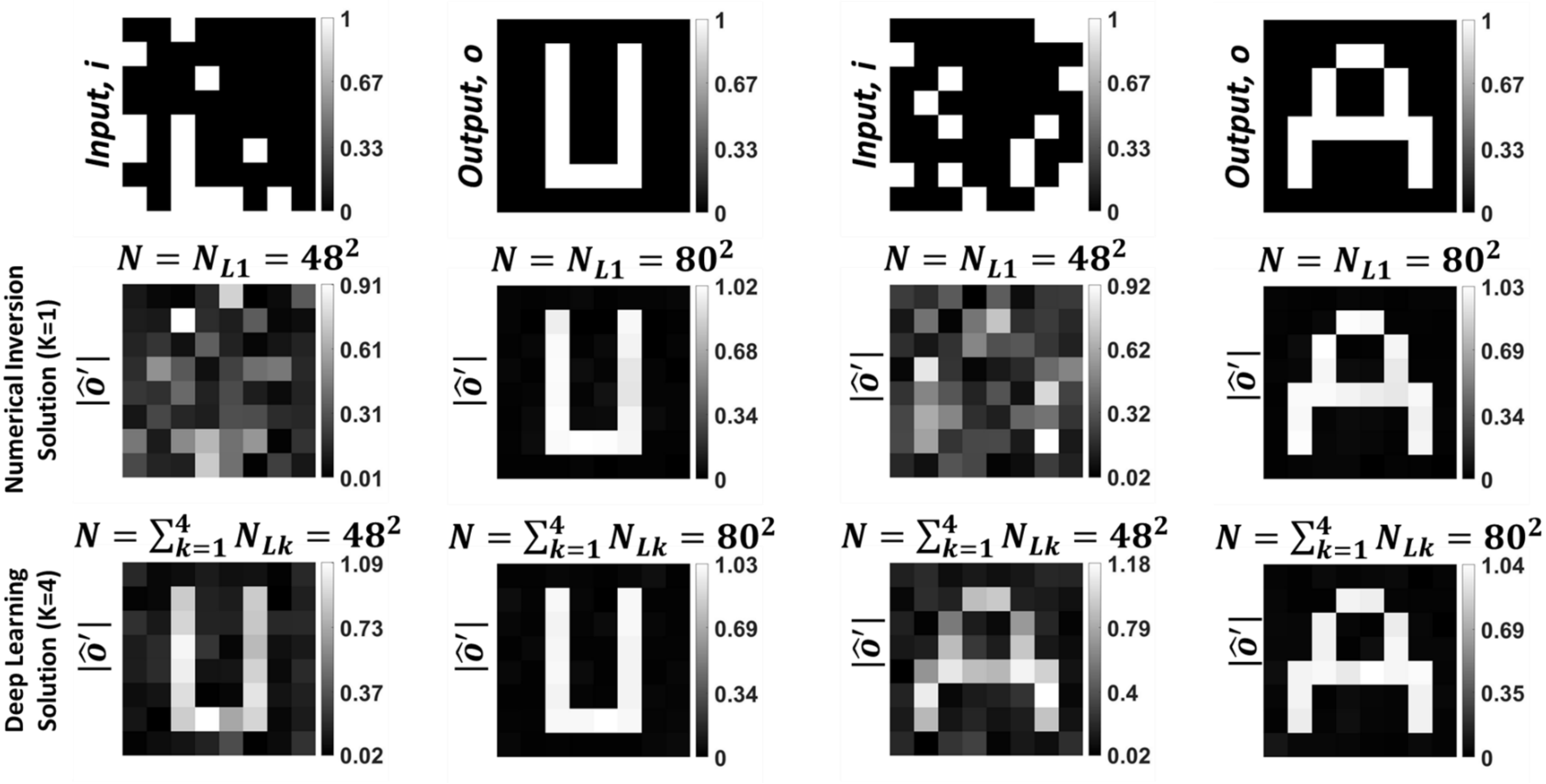

***Figure 12***. *Follows the caption of Fig. 3, except that the target ($\boldsymbol{A} = \mathbf{P}$) is a permutation matrix.*